\documentclass[aps,superscriptaddress,showkeys,floatfix,amsmath,amssymb,twocolumn,showpacs]{revtex4}
\usepackage{graphicx} 
\usepackage{epsfig} 
\usepackage{dcolumn}% Align table columns on decimal point 
\usepackage{bm}% bold math
\newcommand{\Tr}{{\rm Tr}\;}
\newcommand{\si}{{\rm si}}
\newcommand{\be}{\begin{eqnarray}}
\newcommand{\ee}{\end{eqnarray}}
\begin{document}

\title{BCS-BEC crossover of neutron pairs in symmetric and
asymmetric nuclear matter} 

\author{J. Margueron}
\affiliation{ Institut de Physique Nucl\'eaire, Universit\'e
Paris-Sud, IN$_2$P$_3$-CNRS, F-91406 Orsay Cedex, France}
\affiliation{Center for Mathematical Sciences, University of Aizu,
Aizu-Wakamatsu, 965-8580 Fukushima, Japan}
\author{H. Sagawa}
\affiliation{Center for Mathematical Sciences, University of Aizu,
Aizu-Wakamatsu, 965-8580 Fukushima, Japan}
\author{K. Hagino}
\affiliation{Department of Physics, Tohoku University, Sendai, 
e980-8578, Japan}

\date{\today}

\begin{abstract}
We propose new types of density dependent contact pairing
interaction which reproduce the pairing gaps in symmetric and neutron
matter obtained by 
a microscopic treatment based on the nucleon-nucleon interaction.
These interactions are able to simulate the pairing gaps of
either the bare interaction or the
interaction screened by the medium polarization effects.
It is shown that the medium polarization effects cannot be cast into the
density power law function usually introduced together with the contact interaction 
and require the introduction of another isoscalar term.
The BCS-BEC crossover of neutrons pairs in symmetric
  and asymmetric nuclear matter is studied by using these
  contact interactions.
It is shown that the bare and screened pairing interactions
lead to different features of the BCS-BEC crossover in symmetric
nuclear matter. 
For the screened pairing interaction, a two-neutron BEC 
state is formed in symmetric matter at $k_{Fn}\sim 0.2$~fm$^{-1}$
(neutron density $\rho_n/\rho_0\sim 10^{-3}$). 
Contrary the bare interaction does not form the BEC state at
any neutron density.  
\end{abstract}

%effective interactions, nuclear structure models, nuclear matter, intermediate energy heavy-ion reaction, neutron stars


\maketitle

%%%%%%%%%%%%%%%%%%%%%%%%%%%%%%%%%%%%%%%%%%
%%%%%%%%%%%%%%%%%%%%%%%%%%%%%%%%%%%%%%%%%%

\section{Introduction}

Low density nuclear matter tends to be much more interesting than 
a simple zero density limit of the bulk physics of nuclei.
Indeed, new phenomena show up and are mainly associated with the
formation of bound states or with the emergence of strong 
correlations~\cite{bal04,sed03,bro00,bar06,hor05,sed06}.
For instance, in symmetric nuclear matter,
neutrons and protons become strongly correlated 
while the density decreases and
a deuteron BEC state appears
at very low density~\cite{bal95}.
The deuteron-type correlations give extra binding to the nuclear 
equation of state 
and induce new features
at low density~\cite{mar07}.
This transition belongs to 
the BCS-BEC crossover phenomena
which have been extensively studied in several domains of physics and
has recently been
experimentally accessible in cold atomic gas 
(see Ref.~\cite{bul05} and references therein).

In nuclear matter, neutron pairs are also strongly correlated. 
Theoretical predictions suggest that, 
at density around $\rho_0/10$ where $\rho_0$=0.16~fm$^{-3}$,
the $^1$S pairing gap may take a considerably larger value 
than that at normal nuclear density $\rho_0$~\cite{lom99}. 
The density dependence of the pairing gap 
at low density is unfortunately not yet completely clarified and still
awaits a satisfactory solution~\cite{hei00,sch02,cao06}.
Therefore, it could be interesting to explore pairing
interactions based on Brueckner theory
and its consequences to the BCS-BEC crossover. 
Indeed, pairing at low density is relevant for different purposes:
for the understanding of
neutron-rich exotic nuclei near the drip 
line~\cite{ber91,esb92,esb97,dob96,hag05,pil07}, where 
the long tails of density profiles give rise to "halo" or "skin" behavior,
or for the expanding nuclear matter in heavy ion collisions~\cite{cho04},
or even for the physics of neutron stars, where 
several physical phenomena, such as cooling and glitches, 
are thought to depend very sensitively on the size and the
density dependence of the pairing gap~\cite{yak04,sed07,mon07}.
In $^{11}$Li, the wave function of the two neutrons participating to
the halo nucleus has been analyzed with respect to the BCS-BEC
crossover~\cite{hag07}. 
It has been shown that as the distance between the center of mass of
the two neutrons and the core increases, the wave function changes from 
the weak coupling BCS regime to 
the strongly correlated BEC regime.
This is due to the fact that the pairing correlations are
  strongly density-dependent~\cite{mat06} and the distance between the
  two neutrons and the core provides a measure of the pairing strength.

It should be emphasized that the bare nuclear interaction in the
particle-particle channel should be corrected by the medium polarization
effects~\cite{lom99,cao06} (usually referred to as the screening effects).
These effects have been neglected for a long time
since the nuclear interaction is already attractive in the
  $^1$S$_0$ channel without the medium polarization effects,
contrary to the Coulomb interaction for which the medium polarization
effects are absolutely necessary to get an attractive interaction
between electrons. 
However, several many-body methods have been developed recently to
include the medium polarization effects in the calculation of the pairing gap 
such as a group renormalization method~\cite{sch03}, Monte-Carlo methods
calculation~\cite{fab05,abe07,gez07} and extensions of the Brueckner
theory~\cite{lom99,cao06}.
These calculations, except the one presented in Ref.~\cite{fab05},
predict a reduction of the pairing gap %by a factor 2-3 
in neutron matter.% due to the medium polarization effects.

Note that, based on the nuclear field model, it has also been
suggested that the medium polarization effects contribute 
to the pairing interaction in finite nuclei
and increase the pairing gap~\cite{bar99,gio02,bar05}. 
To understand this apparent contradiction between neutron matter and
finite nuclei, a Brueckner calculation
including the medium polarization effects in both symmetric and neutron matter
has been performed in Ref.~\cite{cao06}.
It has been shown that 
the medium polarization effects are different in neutron matter and in
symmetric matter.
The medium polarization effects do not reduce the pairing
gap in symmetric matter, contrary to that in neutron matter.
Instead, in symmetric matter, the neutron pairing gap is 
much enlarged at low density
compared to that of the bare calculation.
This enhancement takes place especially for
neutron Fermi momenta $k_{Fn}<0.7$~fm$^{-1}$.
This could explain why the medium polarization effects increase largely
the pairing correlations in finite nuclei
but decrease it in neutron matter.

In this paper, we propose
an effective density-dependent pairing interaction 
which reproduces both the neutron-neutron (nn) 
scattering length at zero density and the
neutron pairing gap in uniform matter obtained by 
a microscopic treatment based on the nucleon-nucleon interaction~\cite{cao06}.
%the Brueckner G-matrix calculation~\cite{cao06}. 
The proposed interaction has isoscalar and isovector terms which
  could simultaneously describe the density dependence of the neutron
  pairing gap for both symmetric and neutron matter.
Furthermore, we invent different density-dependent
interactions to describe the
``bare'' and ``screened'' pairing gaps, together with the asymmetry of
uniform matter, given in Ref.~\cite{cao06}.
Then, we explore the BCS-BEC crossover phenomena in symmetric and
asymmetric nuclear matter.

This paper is organized as follows.
In Sec.~\ref{sec:int}, we discuss how to determine
the isoscalar and isovector density-dependent contact interactions. 
Applications of those interactions to the BCS-BEC crossover are
presented in Sec.~\ref{sec:crossover}. We give
our conclusions in Sec.~\ref{sec:conclusions}.

%%%%%%%%%%%%%%%%%%%%%%%%%%%%%%%%%%%%%%%%%%
%%%%%%%%%%%%%%%%%%%%%%%%%%%%%%%%%%%%%%%%%%

\section{Density-dependent pairing interaction}
\label{sec:int}

%%%%%%%%%%%%%%%%%%%%%%%%%%%%%%%%%%%%%%%%%%
%%%%%%%%%%%%%%%%%%%%%%%%%%%%%%%%%%%%%%%%%%

Recently, spatial structure of neutron Cooper pair in low density
nuclear matter has been studied using both finite range interactions
like Gogny or G3RS and density-dependent contact interactions properly
adjusted to mimic the pairing gap obtained with the former
interactions~\cite{mat06}. 
It was found that the contact interactions provide almost equivalent
  results compared to the finite range ones for many properties of the Cooper
  pair wave functions.
It is thus reasonable to investigate the evolution of the Cooper pair
wave function with respect to the density and the isospin asymmetry using
contact interactions adjusted to realistic interactions.
%based on G-matrix theory.
In this paper, we take a contact interaction $v_{nn}$ acting on the
singlet $^1$S channel,
\be
\langle k|v_{nn}|k'\rangle=\frac{1-P_\sigma}{2}v_0 \,g[\rho_n,\rho_p] 
\,\theta(k,k') \; ,
\label{eq:pairing_interaction}
\ee
where the cutoff function $\theta(k,k')$ is introduced to remove the ultraviolet
divergence in the particle-particle channel.
A simple regularization could be done by introducing a cutoff 
momentum $k_c$. 
That is, $\theta(k,k')=1$ if $k,k'<k_c$ and 0 otherwise.
In finite systems, a cutoff energy $e_c$ is usually introduced
instead of a cutoff momentum $k_c$.
The relation between the cutoff energy and the cutoff momentum
may depend on the physical problem,
and it is known that
the pairing strength $v_0$ depends strongly on the 
cutoff.
A detailed discussion on the different prescriptions used in
the literature are then presented in Appendix~\ref{app:co}. 
In this paper, we choose the prescription 3 in the Appendix~\ref{app:co}
so that the adjusted interaction can be directly applied to 
Hartree-Fock-Bogoliubov calculations.

In Eq.~(\ref{eq:pairing_interaction}), the interaction strength $v_0$ is 
determined by the low energy scattering phase-shift, that fixes the relation 
between $v_0$ and the cutoff energy $e_c$, 
while the density-dependent term $g[\rho_n,\rho_p]$ is deduced from 
%the G-matrix
predictions of the pairing gaps in symmetric and neutron matter
based on the nucleon-nucleon interaction~\cite{cao06}.
The density-dependent term accounts for the medium 
effects and satisfies the 
boundary condition $g\rightarrow 1$ for $\rho\rightarrow 0$.
The volume type and surface type pairing interactions have $g=1$ and 
$g=0$ at $\rho=\rho_0$, respectively.
In this paper, we introduce
more general types of pairing interactions and the
novelty is 
a dependence on the ratio of neutron to proton
composition of the considered system.
We thus define a function 
\be
g_1[\rho_n,\rho_p] =  1
-f_s(I)\eta_s\left(\frac{\rho}{\rho_0}\right)^{\alpha_s}
-f_n(I)\eta_n\left(\frac{\rho}{\rho_0}\right)^{\alpha_n}
 , \label{eq:ipa}
\ee
where $I$ is the asymmetry parameter, defined as $I=(N-Z)/(N+Z)$, 
and $\rho_0$=0.16~fm$^{-3}$ is the saturation density of symmetric nuclear
matter. 
We insert the function $g_1$ into
Eq.~(\ref{eq:pairing_interaction}) as $g=g_1$.
The goal of the functional form in Eq.~(\ref{eq:ipa}) is to
reproduce the theoretical
calculation of the pairing gap in both symmetric and neutron matter
and also to be used
for prediction of the pairing gap in asymmetric matter.
It could also be applied to describe pairing correlations in finite 
nuclei by acquiring an explicit dependence on the 
coordinate $r$ in the density $\rho(r)$ 
and the asymmetry parameter $I(r)$.
In Eq.~(\ref{eq:ipa}), the interpolation functions $f_s(I)$ and
$f_n(I)$ are not explicitly known but should satisfy the following
condition $f_s(0)=f_n(1)=1$ and $f_s(1)=f_n(0)=0$. 
The density-dependent function $g_1$ is flexible enough and we 
can obtain an effective pairing interaction which reproduces 
the density dependence of the pairing gap 
%obtained by G-matrix calculations
in uniform matter.
It should however be noticed that the interpolation functions $f_s(I)$
and $f_n(I)$ cannot be deduced from the adjustment of the pairing gap
in symmetric and neutron matter.
For that, theoretical calculations in asymmetric nuclear matter or
application to exotic nuclei are necessary.
In this paper, we 
choose $f_s(I)=1-f_n(I)$ and $f_n(I)=I$.
Many different interpolation functions could be explored but we think
that it has little consequences to the BCS-BEC crossover.

We have also explored other density-dependent functionals,
introducing an explicit dependence on the isovector density,
$1-\eta_{s}\left(\frac{\rho}{\rho_0}\right)^{\alpha_{s}}
-\eta_{i}\left(\frac{\rho_n-\rho_p}{\rho_0}\right)^{\alpha_{i}}$,
or introducing the neutron and proton densities,
$1-\eta_n\left(\frac{\rho_n}{\rho_0}\right)^{\alpha_n}
-\eta_p\left(\frac{\rho_p}{\rho_0}\right)^{\alpha_p}$.
The isospin dependence of those functionals is fixed
but these functional forms are not flexible enough so that
the adjustment becomes sometimes impossible.
For instance, the pairing gap in symmetric and neutron matter including 
the medium polarization effects and calculated in Ref.~\cite{cao06} 
cannot be reproduced with such functionals.
In the following, we have therefore 
considered only the density-dependent interaction
given in Eq.~(\ref{eq:ipa}).

Recently, another attempt has been done to introduce an isospin dependent
volume type pairing interaction in a very different approach~\cite{gor04}.
This interaction has been adjusted to reproduce empirical
mass formula over a thousands of
nuclei, but the pairing gap calculated with
this pairing interaction compares very badly with realistic
%G-matrix
calculations in uniform matter.

\subsection{The free interaction}
\label{ssec:fi}

%%%%%%%%%%%%%
%%%%%%%%%%%%%
%%%%%%%%%%%%%
\begin{figure}[tb]
\begin{center}
\includegraphics[scale=0.33]{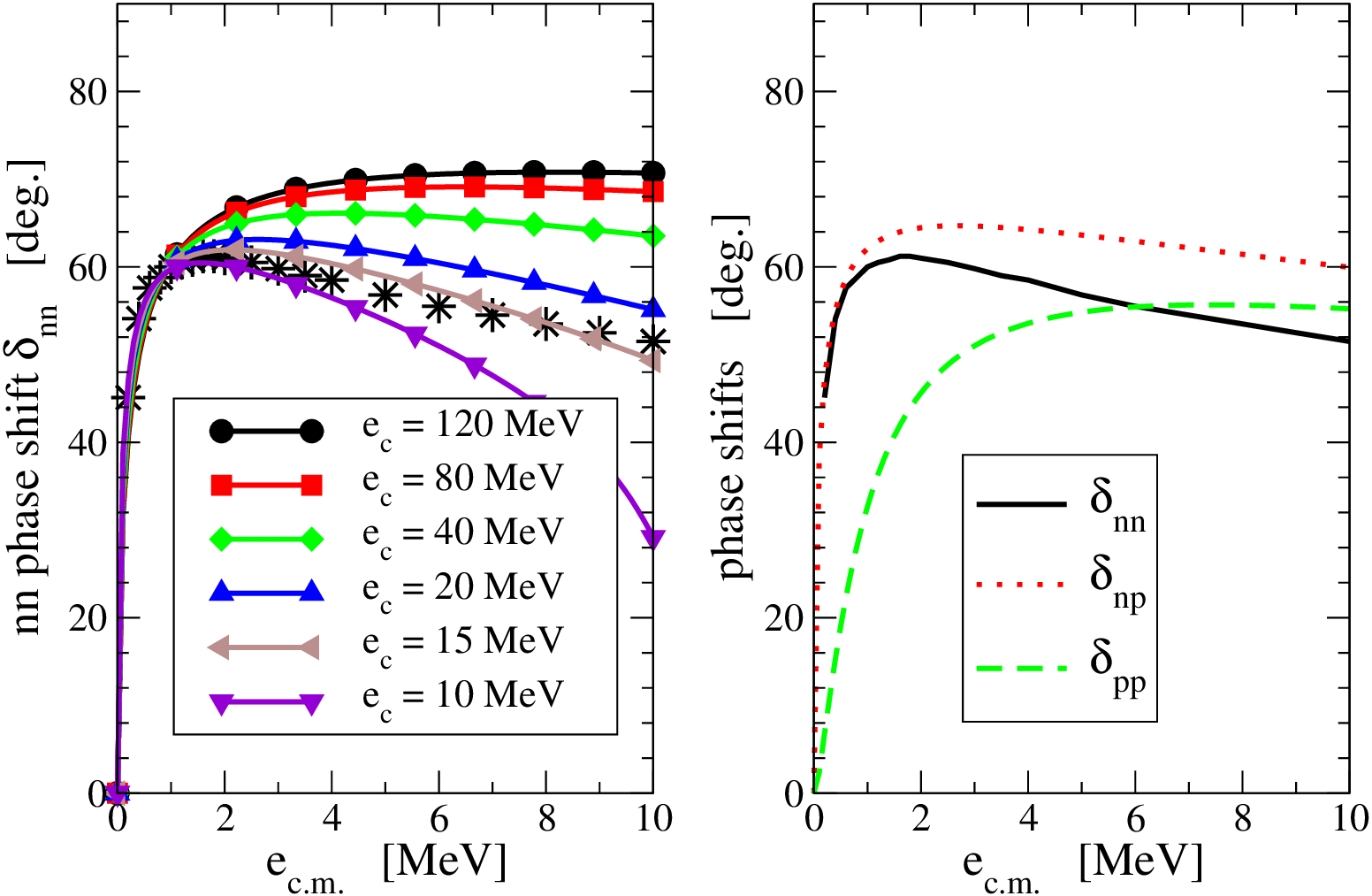}
\end{center}
\caption{
(Color online) 
Phase shifts for s-wave nucleon-nucleon 
scattering as a function of the center of mass
energy. In the left panel are shown nn phase shifts 
obtained from Argonne $v_{18}$ potential~\cite{esb97} (stars) and 
the result of the best adjustment obtained with a contact interaction 
for a set of cutoff energies (from 10 to 120~MeV). 
In the right panel are shown the s-wave phase shifts for various
channels: nn, np and pp. The np and pp phase shifts have been provided 
by the Nijmegen group (http://nn-online.org).}
\label{fig01}
\end{figure}
%%%%%%%%%%%%%
% Obtenu avec delta1.gnu
%%%%%%%%%%%%%
%%%%%%%%%%%%%
In Ref.~\cite{ber91}, it has been proposed to deduce the free interaction
parameter $v_0$ from the low energy phase shift of nucleon-nucleon scattering.
The nn, np and pp phase shifts versus the center of mass energy are shown
on the right panel of Fig.~\ref{fig01}. It is clear from this figure that each
of these three channels are different and the interaction parameter
$v_0$ should depend on the channel of interest. 
In this paper, we are interested only in the nn channel.
We then express the phase shift as a function of the cutoff momentum
$k_c$ and the scattering length $a_{nn}$~\cite{esb97,gar99}: 
\be
k \cot \delta&=&-\frac{2}{\alpha \pi}
\left[1+\alpha k_c+\frac{\alpha k}{2}\ln\frac{k_c-k}{k_c+k}
\right]\\
&=& -\frac{1}{a_{nn}} - \frac{k}{\pi}\ln\frac{k_c-k}{k_c+k}
\ee
where $\alpha=2a_{nn}/(\pi-2k_c a_{nn})$.
In this way, for a given cutoff momentum $k_c$, the phase shift 
can be adjusted using the scattering length $a_{nn}$ as a
variable.
%%%%%%%%%%%%%%%%%%%%%%%%
\begin{table}[tb]
\begin{center}
\renewcommand{\arraystretch}{1.3}
\begin{tabular}{ccccccc}
\toprule 
 $e_c$ & $a_{nn}$ & $r_{nn}$ & $\alpha$ & $v_0$ & $v_0^*$ & $v_0^\infty$ \\
 \,[MeV] & [fm] & [fm] & [fm] & [MeV.fm$^3$] & [MeV.fm$^3$] & [MeV.fm$^3$]  \\
\colrule
120 & $-$12.6 & 0.75 & $-$0.55 & $-$448  & $-$458  & $-$481 \\
80  & $-$13.0 & 0.92 & $-$0.66 & $-$542  & $-$555  & $-$589 \\
40  & $-$13.7 & 1.30 & $-$0.91 & $-$746  & $-$767  & $-$833 \\
20  & $-$15.0 & 1.83 & $-$1.25 & $-$1024 & $-$1050 & $-$1178\\
15  & $-$15.7 & 2.12 & $-$1.43 & $-$1167 & $-$1192 & $-$1360\\
10  & $-$17.1 & 2.59 & $-$1.72 & $-$1404 & $-$1421 & $-$1666\\
\botrule
\end{tabular}
\end{center}
\caption{For a given cutoff energy $e_c$, the
parameters $r_{nn}$, $\alpha$ and $v_0$ are determined by the
scattering length $a_{nn}$ which reproduces the phase shift
in the low energy region $e_{c.m.}<2$~MeV.
The interaction strengths $v_0^*$ and $v_0^\infty$ are obtained from the 
empirical value of $a_{nn}$=-18.5~fm and from the unitary limit, 
defined as $a_{nn}\rightarrow\infty$.}
\label{tab1}
\end{table}%
%%%%%%%%%%%%%
% Obtenu avec delta1.gnu
%%%%%%%%%%%%%
%%%%%%%%%%%%%
The results are shown in Fig.~\ref{fig01} and
Table~\ref{tab1} for a set of cutoff energies $e_c=\hbar^2k_c^2/m$
(note that we use the reduced mass $m/2$).
We have found that, for cutoff energies larger than 20~MeV, 
the optimal parameters cannot reproduce
the nn phase shift in the
low energy region
($e_{c.m.}<$2~MeV) and in the higher energy region simultaneously. 
We mainly focus on the adjustment of the nn phase shift in the
low energy region. 
At higher energies, or equivalently at higher densities in nuclear
matter, the medium effects modify the interaction anyhow and generate a density
dependent term in the interaction.

Fixing $e_c$ and $a_{nn}$, one can deduce the effective range 
$r_{nn}=4/\pi k_c$, the parameter 
$\alpha$ and the interaction strength 
$v_0=2\pi^2\alpha\hbar^2/m$.
The values of those parameters are given in Table~\ref{tab1}.
The value of the free interaction parameter $v_0^*$ deduced from the 
empirical value of the scattering length $a_{nn}=-18.5$~fm is also indicated.
One can see that the difference between $v_0$ and $v_0^*$ is
small, as much as 3\%.
Indeed, as we are in a regime of large scattering length, 
one can deduce the interaction strength approximately from the relation
$v_0\approx v_0^\infty\left(1+\pi/2k_ca_{nn}+\dots\right)$ where
$v_0^\infty=-2\pi^2\hbar^2/mk_c$ is the interacting strength in the
unitary limit ($k_c a_{nn}\rightarrow\infty$).

\subsection{The density-dependent function $g[\rho_n,\rho_p]$}
\label{ssec:dd}

The density-dependent function $g$ is adjusted to reproduce the pairing gaps
in symmetric and neutron matter obtained from Ref.~\cite{cao06}.
Pairing in uniform nuclear matter is evaluated with the BCS ansatz:
\be
|BCS\rangle=\prod_{k>0} (u_k+v_k \hat{a}^\dagger_{k\uparrow}
\hat{a}^\dagger_{-k\downarrow})|-\rangle \; , 
\label{eq:bcs}
\ee
where $u_k$ and $v_k$
represent the BCS variational parameters and $\hat{a}^\dagger_{k\uparrow}$ are
creation operators of a particle with momentum $k$ and spin
$\uparrow$ on top of the vacuum $|-\rangle$~\cite{bcs57,gennes,schuck}.
The BCS equations are deduced from the minimization of the energy with
respect to the variational parameters $u_k$ and $v_k$.
For a contact interaction, the equation for the pairing gap
$\Delta_n$ takes the following simple form at zero temperature,
\be
\Delta_n = -\frac{v_0 g[\rho_n,\rho_p]}{2(2\pi)^3} \int d^3 k 
\frac{\Delta_n}{E_n(k)} \theta(k,k) \;,
\label{eq:gap}
\ee
where $\theta(k,k)$ is the cutoff function associated to the contact
interaction~(\ref{eq:ipa}),
$E_n(k)=\sqrt{(\epsilon_n(k)-\nu_n)^2+\Delta_n^2}$ is the 
neutron quasi-particle energy, $\epsilon_n(k)=\hbar^2k^2/2m_n^*$ 
is the neutron single particle kinetic energy 
with the effective mass $m^*_n$.
We define the effective neutron chemical potential $\nu_n=\mu_n-U_n$,
where the neutron mean field potential $U_n$ is subtracted
from the neutron chemical potential $\mu_n$. 
The effective neutron chemical potential $\nu_n$ gives
the neutron density, 
\be
\rho_n=\frac{2}{V}\sum_k n_n(k)
\label{eq:rho}
\ee
where $V$ is the volume and $n_n(k)$ is the occupation probability
defined as
\be
n_n(k)= \frac{1}{2}\left[ 1-\frac{\epsilon_n(k)-\nu_n}{E_n(k)}\right]
\; .
\label{eq:nn}
\ee
Finally, the neutron Fermi momentum $k_{Fn}$ is defined as 
$\rho_n\equiv k_{Fn}^3/3\pi^2$.

\begin{figure}[tb]
\begin{center}
\includegraphics[scale=0.35]{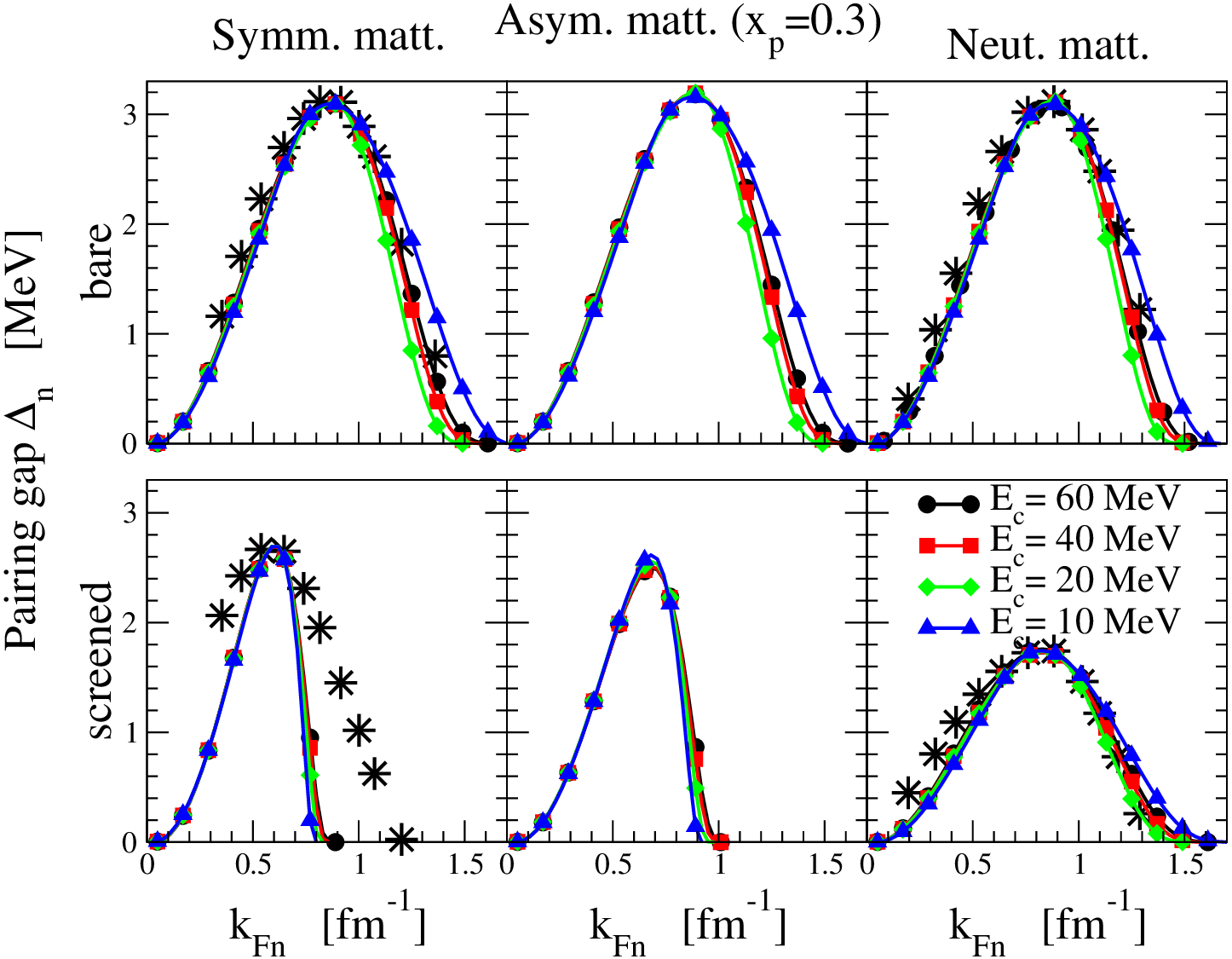}
\end{center}
\caption{(Color online) 
Pairing gap in symmetric, asymmetric ($x_p=\rho_p/\rho$) and neutron matter 
adjusted to the "bare gap" (upper panel) or to the "screened gap" (lower panel)
for various cutoff energies $E_c$. 
The pairing gap calculated from the microscopic treatment presented 
%Brueckner G-matrix theory 
in Ref.~\cite{cao06} is also shown as star symbols.}  
\label{fig02}
\end{figure}

%%%%%%%%%%%%%%%%%%%%%%%%
\begin{table}[tb]
\begin{center}
\setlength{\tabcolsep}{.08in}
\renewcommand{\arraystretch}{1.3}
\begin{tabular}{cccccc}
\toprule 
& $E_c=e_c/2$ & $\eta_s$ & $\alpha_s$ & $\eta_n$ & $\alpha_n$ \\
\colrule
\hbox{bare} & 60~MeV & 0.598 & 0.551 & 0.947 & 0.554 \\
$g=g_1$      & 40~MeV & 0.664 & 0.522 & 1.01  & 0.525 \\
            & 20~MeV & 0.755 & 0.480 & 1.10  & 0.485 \\
            & 10~MeV & 0.677 & 0.365 & 0.931 & 0.378 \\
\hline
\hbox{screened-I} & 60~MeV & 7.84 & 1.75 & 0.89 & 0.380 \\
$g=g_1$  & 40~MeV & 8.09 & 1.69 & 0.94 & 0.350 \\
                & 20~MeV & 9.74 & 1.68 & 1.00 & 0.312 \\
                & 10~MeV & 14.6 & 1.80 & 0.92 & 0.230 \\
\hline
\hbox{screened-II} & 60~MeV & 1.61 & 0.23 & 1.56 & 0.125 \\
$g=g_1+g_2$      & 40~MeV & 1.80 & 0.27 & 1.61 & 0.122 \\
$\eta_2=0.8$       & 20~MeV & 2.06 & 0.31 & 1.70 & 0.122 \\
                 & 10~MeV & 2.44 & 0.37 & 1.66 & 0.0939 \\
\botrule
\end{tabular}
\end{center}
\caption{Parameters of the function $g$
defined in Eq.~(\ref{eq:ipa}).
These parameters are obtained from the
  fit to the pairing gaps in symmetric and neutron matter.
  These are the parameters obtained from the adjustment of the bare
  gap and the screened gap with $g=g_1$, and the screened gap 
  including the additional function $g_{2}$.
  The effective mass is obtained from SLy4 Skyrme interaction.
  Note that $E_c$ is the cutoff for the quasi-particle gap
  equation~(\ref{eq:gap}) while $e_c$ is that for the two-body
  scattering so that $E_c=e_c/2$. See the text for details.}
\label{tab2}
\end{table}%
%%%%%%%%%%%%%
% Obtenu avec gap.f
%%%%%%%%%%%%%
%%%%%%%%%%%%%
We have chosen to adjust our interaction to the results of
nuclear matter pairing gaps in
Ref.~\cite{cao06} since it is the only calculations performed for both
symmetric and neutron matter.
We adjust the contact pairing interaction so that it reproduces the
position and the absolute values of the maxima of the pairing gaps in 
symmetric and neutron matter.
For the bare pairing gap, the maximum is located at $k_{Fn}=0.87$~fm$^{-1}$
with $\Delta_n$=3.1~MeV for both symmetric and neutron matter, 
while for the screened pairing gap, the maximum
is at $k_{Fn}=0.60$~fm$^{-1}$ with $\Delta_n$=2.70~MeV for symmetric matter and
$k_{Fn}=0.83$~fm$^{-1}$ and $\Delta_n$=1.76~MeV for neutron matter. 
The value of the parameters $\eta_s$ and $\eta_n$
are freely explored in the real axis while the parameters $\alpha_s$ and
$\alpha_n$ are imposed to be positive to avoid singularities.
The neutron effective mass $m_n^*$ is obtained from SLy4 Skyrme 
interaction since it is widely used in nuclear mean-field calculations.
The results of the fits are given in Table~\ref{tab2} and the pairing
gaps are shown in Fig.~\ref{fig02}.
%%%%%%%%%%%%%
%%%%%%%%%%%%%
%%%%%%%%%%%%%
%%%%%%%%%%%%%
% Obtenu avec hfbcs3.f
%%%%%%%%%%%%%
%%%%%%%%%%%%%

One should note that for the bare interaction, even if the pairing gap
is identical in symmetric and neutron matter, the adjusted contact interaction
is not necessarily isoscalar. Indeed, the transformation from 
the Fermi momentum, the x-axis of Fig.~\ref{fig02}, to the density
is different in symmetric nuclear matter, $\rho/\rho_0=(k_{Fn}/k_{F0})^3$ 
(where $\rho_0=2/(3\pi^2) k_{F0}^3$=0.16~fm$^{-3}$), 
and in neutron matter, $\rho/\rho_0=0.5(k_{Fn}/k_{F0})^3$.
Therefore, an interaction which depends only on the ratio $\rho/\rho_0$
gives different results if it is plotted as a function of $k_{Fn}$
in symmetric and neutron matter.
As the pairing gap calculated with the bare interaction~\cite{cao06}
is quasi-identical in symmetric and neutron
matter when it is plotted versus $k_{Fn}$, one can then deduce the
following relations between the parameters of the density-dependent
term $g_1$ (neglecting the isospin dependence of the effective mass):
$\alpha_s=\alpha_n$ and $\eta_s=\eta_n/2^{\alpha_n}$. 

\begin{figure}[tb]
\begin{center}
\includegraphics[scale=0.36]{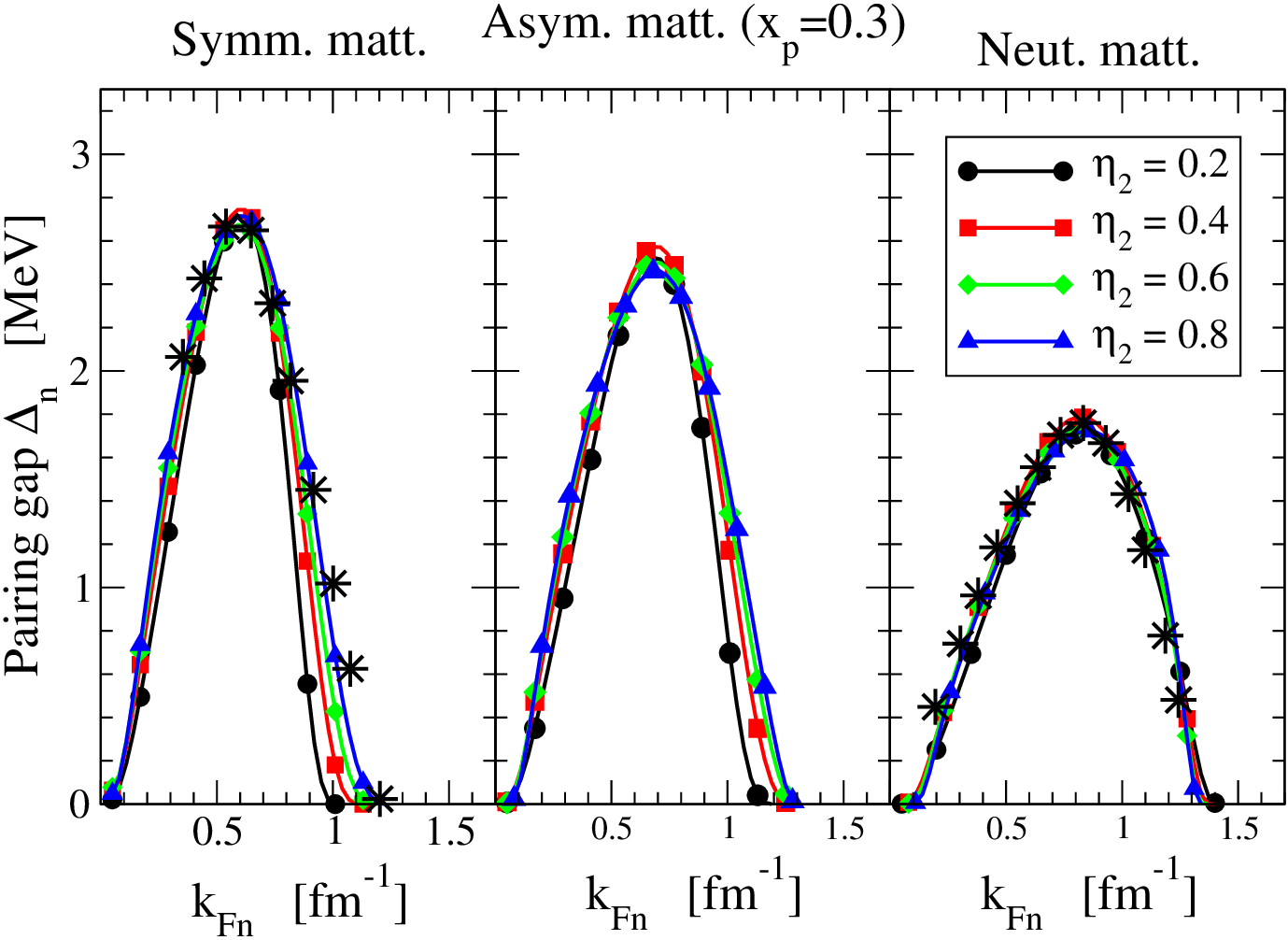}
\end{center}
\caption{(Color online) Pairing gap calculated in symmetric, asymmetric and neutron matter 
with the screened-II interaction ($g=g_1+g_2$)
and for several values of $\eta_{2}$ as indicated in the 
legend. The corresponding parameters are given in
Table~\ref{tab4}. See the text for details.}
\label{fig03}
\end{figure}

For the bare pairing gap and
for a given cutoff energy $E_c$, the position and the maximum 
value of the gap are reproduced well by the contact interaction
in Fig.~\ref{fig02}.
However, in the high Fermi momentum region $k_{Fn}>1$~fm$^{-1}$,
we can see appreciable difference between the microscopic predictions
%G-matrix 
and the pairing gap obtained from the contact interactions.
The best agreement is obtained for a cutoff energy $E_c=40$~MeV.
In the screened case,
the dependence of the pairing gap on $k_{Fn}$ is badly reproduced,
especially for symmetric nuclear matter.
This is because the maximum position
of the pairing gap is shifted towards a lower neutron Fermi momentum
(one third  in density from that for the bare gap).
Consequently, the density dependence of the function $g_1$ becomes stiffer
in the ``screened'' case than in the bare case, 
and the gap drops faster after the maximum, as it is shown
in Fig.~\ref{fig02}.
This may indicate that the screened interaction has a different
density dependence and cannot be cast into a simple power law of the
density as in Eq.~(\ref{eq:ipa}).
Indeed, in Ref.~\cite{cao06}, the medium polarization effects have been
analyzed at the level of the interacting potential, and it was shown that 
the medium polarization effects emerge
at very low density and remain relatively constant.
To simulate such effects, it seems necessary to introduce a new term, 
$g_2$ in Eq.~(\ref{eq:ipa}).
We propose for $g_2$ a simple isoscalar constant 
which switches on at a very low value of the density 
($k_F\sim0.15$~fm$^{-1}$) and switch off around the saturation density.
The following form satisfies this condition, 
\be
g_{2}=\eta_{2}\left[\left(1+e^{\frac{k_F-1.4}{0.05}}\right)^{-1}-
\left(1+e^{\frac{k_F-0.1}{0.05}}\right)^{-1}\right]
\; .
\label{eq:g2}
\ee
The new pairing interaction with $g=g_1+g_2$, hereafter named screened-II, 
has then only one new adjustable parameter $\eta_{2}$.
As the medium polarization effects could also change the density-dependent
term $g_1$, all the 5 parameters have to be re-adjusted.
In Tables~\ref{tab2} and \ref{tab4}, we give the new parameters 
$\eta_s$, $\alpha_s$, $\eta_n$ 
and $\alpha_n$, obtained for several values of $\eta_{2}$. 
The cutoff energy is fixed to be $E_c$=40~MeV. 
The corresponding pairing gap is represented in Fig.~\ref{fig03}.
%%%%%%%%%%%%%
%%%%%%%%%%%%%
%%%%%%%%%%%%%
%%%%%%%%%%%%%
% Obtenu avec gap.f
%%%%%%%%%%%%%
%%%%%%%%%%%%%
The best fit is obtained for the value $\eta_{2}$=0.8.
Eq.~(\ref{eq:g2}) may not be a unique way to take into account the
medium polarization effects. Nevertheless it is simple enough to apply to the
BCS-BEC crossover, so this is why we adopt this functional form.
%%%%%%%%%%%%%%%%%%%%%%%%
\begin{table}[b]
\begin{center}
\setlength{\tabcolsep}{.15in}
\renewcommand{\arraystretch}{1.3}
\begin{tabular}{ccccc}
\toprule 
$\eta_{2}$ & $\eta_s$ & $\alpha_s$ & $\eta_n$ & $\alpha_n$ \\
\colrule
0.2 & 1.90 & 0.72 & 1.08 & 0.24 \\
0.4 & 1.61 & 0.46 & 1.26 & 0.19 \\
0.6 & 1.64 & 0.33 & 1.44 & 0.15 \\
0.8 & 1.80 & 0.27 & 1.61 & 0.122 \\
\botrule
\end{tabular}
\end{center}
\caption{Parameters of the the screened-II interaction, 
where the density-dependent function of the pairing interaction is 
taken to be $g=g_1+g_2$. The functional forms $g_1$ and
$g_2$ are obtained by fitting the screened pairing gap for several values of $\eta_{2}$. 
The energy cutoff is taken to be $E_c=40$~MeV and the neutron 
effective mass is deduced from the SLy4 Skyrme interaction.}
\label{tab4}
\end{table}%
%%%%%%%%%%%%%
% Obtenu avec gap.f
%%%%%%%%%%%%%
%%%%%%%%%%%%%

\begin{figure}[tb]
\begin{center}
\includegraphics[scale=0.36]{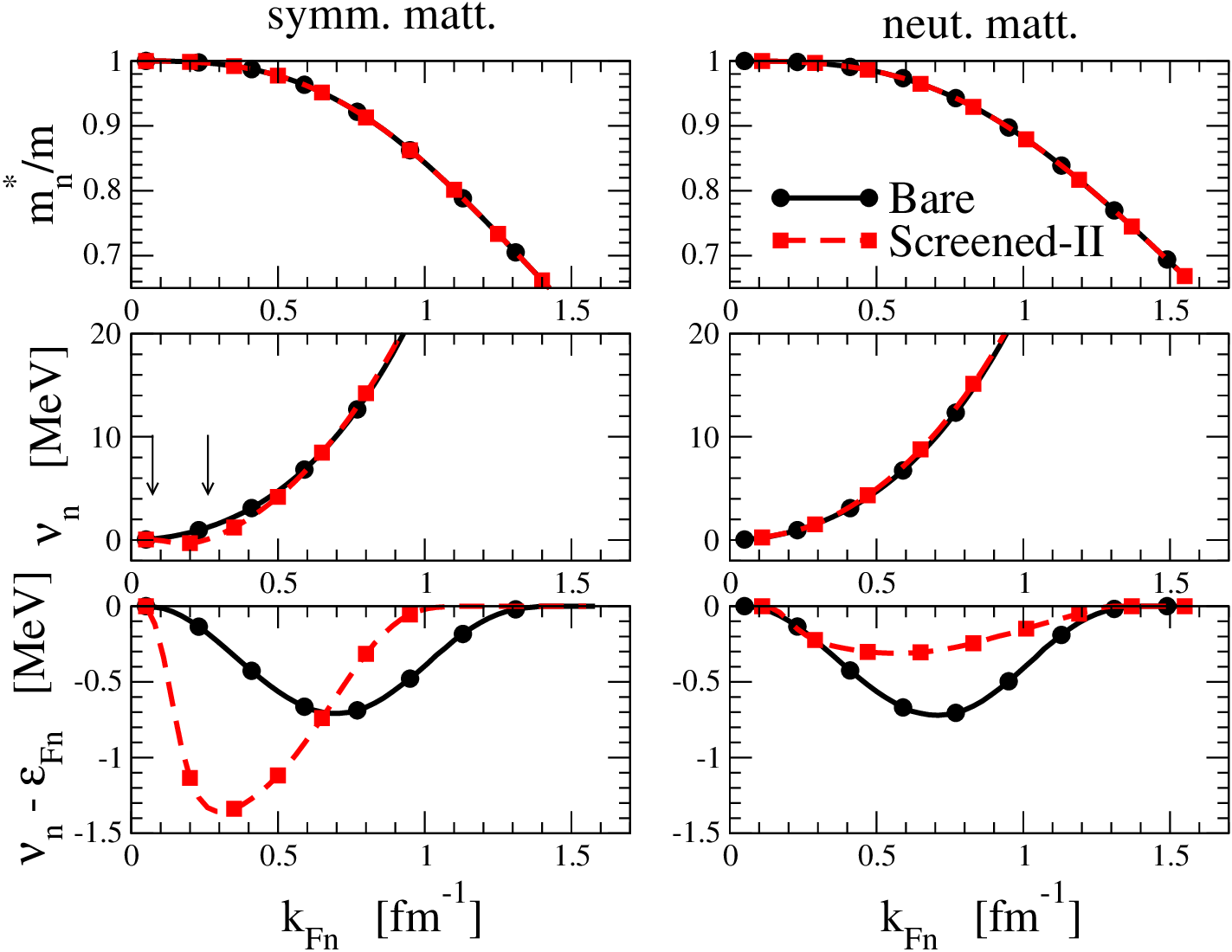}
\end{center}
\caption{(Color online) Comparison of 
the neutron effective mass $m^*_n$, the
effective chemical potential $\nu_n=\mu_n-U_n$, and the difference
$\nu_n-\epsilon_{Fn}$ calculated with the bare and the screened-II
contact interactions. 
The parameters of the pairing interactions
are taken from
Table~\ref{tab2} with the cutoff energy $E_c$=40~MeV.
The effective mass $m_n^*$ and the neutron potential $U_n$ are taken
from SLy4 Skyrme interaction.
The two arrows indicate the lower and upper limits for the
  condition $\nu_n<0$ with the screened-II interaction in symmetric
  nuclear matter.}
\label{fig04}
\end{figure}

Solving the gap Eq.~(\ref{eq:gap}), the neutron effective
chemical potential $\nu_n$ is determined for a given interaction at a
given neutron Fermi momentum $k_{Fn}$.
The neutron effective mass $m^*_n$, 
the effective neutron chemical potential $\nu_n=\mu_n-U_n$ and 
the difference $\nu_n-\epsilon_{Fn}$ are represented in Fig.~\ref{fig04}
as a function of the neutron Fermi momentum in symmetric and neutron matter. 
%%%%%%%%%%%%%
%%%%%%%%%%%%%
%%%%%%%%%%%%%
%%%%%%%%%%%%%
% Obtenu avec gap.f
%%%%%%%%%%%%%
%%%%%%%%%%%%%
The neutron effective mass and the neutron potential $U_n$ are
deduced from the SLy4 Skyrme interaction. 
Note that the neutron density $\rho_n$ is changed into the neutron Fermi 
momentum, $k_{Fn}$. 
We have selected from Table~\ref{tab2} 
the bare and the screened-II pairing
interactions for a cutoff energy $E_c=40$~MeV.
In the absence of pairing, the effective neutron chemical potential
$\nu_n$ is identical to the neutron Fermi kinetic energy,
$\nu_n=\epsilon_{Fn}$ where $\epsilon_{Fn}=\epsilon_n(k_{Fn})$.
The difference $\nu_n-\epsilon_{Fn}$ is then null in the absence of
pairing correlations, otherwise it is negative as shown in
Fig.~\ref{fig04}. 
From this difference, one can estimate the relative importance of the
pairing correlations: in neutron matter the screened-II interaction leads to
weaker pairing correlations compared to the bare one, while in symmetric
matter, the screened-II interaction give much stronger pairing
correlations for $k_{Fn}<0.7$~fm$^{-1}$ and less for
  $k_{Fn}>0.7$~fm$^{-1}$.

It is easy to show that the gap~(\ref{eq:gap}) and the occupation 
probability~(\ref{eq:nn}) go over into the Schr\"odinger-like equation for the
neutron pair wave function $\Psi_{pair}$~\cite{noz85},
\be
\frac{p^2}{m}\Psi_{pair}+[1-2n_n(k)] \frac{1}{V} \Tr v_{nn}\Psi_{pair}
=2\nu_n\Psi_{pair}. 
\label{eq:sch}
\ee
See Eq.(~\ref{eq:psi}) in Sec.~\ref{sec:crossover} for a proper
definition of the neutron pair wave function $\Psi_{pair}$.
Notice that, at zero density where $n_n(k)=0$, 
Eq.~(\ref{eq:sch}) is nothing but the Schr\"odinger equation for the neutron pair.
From Eq.~(\ref{eq:sch}), one usually relates
the effective neutron
chemical potential $\nu_n$ to be a half of the ``binding energy'' of a Cooper pair.
The Cooper pairs may then be considered to be strongly correlated
if $\nu_n$ is negative.
Notice from Fig.~\ref{fig04}, that the effective neutron chemical
potential $\nu_n$ becomes negative with the screened-II interaction
in symmetric nuclear matter for $k_{Fn}$=0.05-0.35~fm$^{-1}$,  
but not at all in neutron matter.
One could then expect in this case that
the Cooper pair may
resemble a closely bound system (BEC) in symmetric nuclear matter at very low density 
while in neutron matter it should behave like that of the weak
coupling BCS regime.
However, to go beyond this rough interpretation, we need to study more
accurately the BCS-BEC crossover in asymmetric matter.

%%%%%%%%%%%%%%%%%%%%%%%%%%%%%%%%%%%%%%%%%%
%%%%%%%%%%%%%%%%%%%%%%%%%%%%%%%%%%%%%%%%%%

\section{Application to the BCS-BEC crossover}
\label{sec:crossover}

Going from the weak coupling BCS regime, around the saturation density
$\rho_0$, down to the BCS-BEC crossover, for densities $<\rho_0/10$, 
it has been shown that the spatial structure of the neutron Cooper 
pair changes~\cite{mat06}.
It is indeed expected that the correlations between two neutrons get
large as the density decreases
and as a consequence, the BCS-BEC crossover 
occurs in the uniform matter at low density. 
However, being of the second order, this transition is smooth.
Hereafter, we clarify the boundaries of the BCS-BEC phase transition
by using a regularized gap equation.

Although the BCS ansatz~(\ref{eq:bcs})
has been developed to describe the Cooper
pair formation in the weak BCS regime~\cite{bcs57}, it has been shown
that the BCS equations are also valid in the strong BEC condensation
regime~\cite{noz85,PS03}. 
The BCS equations are thus 
adopted as a useful framework to
describe the intermediate BCS-BEC crossover
regime at zero temperature~\cite{eng97}. 
It has been proposed to define the limit of the BCS-BEC phase
transition using a regularized model for the pairing
gap~\cite{eng97,pap99,mat06}.  
In this model, the BCS gap~(\ref{eq:gap}) is combined 
with the relation between the interaction strength and the 
scattering length
which has a similar divergent behavior.
The difference between those two divergent integrals
gives a regularized equation,
\be
\frac{m_n}{4\pi a_{nn}}=-\frac{1}{2V} \Tr \left(
\frac{1}{E_n(k)}-\frac{1}{\epsilon_n(k)} \right) \; ,
\label{eq:reg}
\ee
which has no divergence.
The gap equation~(\ref{eq:gap}) can be solved analytically for
the contact interaction with a constraint of the particle number 
conservation~(\ref{eq:rho}).
The solution of this regularized gap equation is
independent of the strength of the interaction, and the gap is
uniquely determined by the value of the scattering length $a_{nn}$.
From Eq.~(\ref{eq:reg}), one can study the
boundaries of the BCS-BEC phase transition with respect to the
dimensionless order parameter $k_{Fn} a_{nn}$. 
%%%%%%%%%%%%%%%%%%%%%%%%%%%%%%%%%%%%%%%%%%
%%%%%%%%%%%%%%%%%%%%%%%%%%%%%%%%%%%%%%%%%%
%%%%%%%%%%%%%%%%%%%%%%%%
\begin{table}[tb]
\begin{center}
\setlength{\tabcolsep}{0.02in}
\renewcommand{\arraystretch}{1.3}
\begin{tabular}{cccccc}
\toprule
$(k_{Fn} a_{nn})^{-1}$ & $P(d_n)$ & \,$\xi_{rms}/d_n$ & \,$\Delta_n/\epsilon_{Fn}$ & \,$\nu_n/\epsilon_{Fn}$  & \\
\colrule
$-$1 & 0.81 & 1.10 & 0.21 & 0.97 & \hbox{BCS boundary}\\
 0 & 0.99 & 0.36 & 0.69 & 0.60 & \hbox{unitarity limit}\\
 1 & 1.00 & 0.19 & 1.33 & $-$0.77& \hbox{BEC boundary}\\
\botrule
\end{tabular}
\end{center}
\caption{Reference values 
of $(k_{Fn} a_{nn})^{-1}$, $P(d_n)$,
$\xi_{rms}/d_n$, $\Delta_n/\epsilon_{Fn}$ and $\nu_n/\epsilon_{Fn}$
  characterizing the BCS-BEC crossover in the regularized 
model for the contact interaction. 
The values $d_n$, $P(d_n)$, and $\xi_{\rm rms}$ are 
the average distance between neutrons $d_n=\rho_n^{-1/3}$, 
the probability for the
partner neutrons 
correlated within the relative distance $d_n$, 
and the rms radius of Cooper pair, respectively. 
The numbers have been taken from Refs.~\cite{eng97,mat06}.
See the text for details.}
\label{tab5}
\end{table}%
%%%%%%%%%%%%%
% Extracted from eng97 and mat06
%%%%%%%%%%%%%
%%%%%%%%%%%%%
We give in Table~\ref{tab5} the values of several quantities which
specify the phase transition: the probability $P(d_n)$ for the
partner neutrons to be correlated within the relative distance $d_n$
($d_n$ is the average distance between neutrons $d_n=\rho_n^{-1/3}$),
the ratio of the rms radius 
to the mean neutron distance
$\xi_{rms}/d_n$, the ratio of the pairing gap to the single particle
kinetic energy $\Delta_n/\epsilon_{Fn}$ and also the ratio of
the effective neutron chemical potential to the single particle
kinetic energy $\nu_n/\epsilon_{Fn}$. 
As we already mentioned, 
these boundaries are 
indicative because the phase transition is smooth
at the boundaries being of the second order. 
For instance, even if the nuclear matter does not enter into the 
BEC regime,
we will show that the Cooper pair wave function is already very
similar to the BEC one when it is close.

A drawback of this regularized model is that the relation between
the dimensionless order parameter $k_{Fn}a_{nn}$ and the density of the
medium is unknown.
To relate the order parameter to the density, one has to 
re-introduce the pairing strength in the gap equation~(\ref{eq:gap}).
We could consider for instance a contact interaction with a cutoff
regularization.
The density will then trigger the phase transition for a given pairing
interaction strength.
In the following, we study the BCS-BEC phase diagram in
asymmetric nuclear matter for the two pairing interactions
discussed in Sec.~\ref{sec:int}. 
Namely we explore the properties of the Cooper pair wave function
obtained by
the bare and the screened-II interactions presented in
Table~\ref{tab2} for a fixed cutoff energy $E_c$=40~MeV.

The BCS approximation provides the Cooper pair wave function 
$\Psi_{pair}(k)$~\cite{bcs57,gennes,schuck}
\be
\Psi_{pair}(k) &\equiv& C \langle BCS|\hat{a}^\dagger(k\uparrow) 
\hat{a}^\dagger(-k\downarrow)|BCS\rangle \\
&\equiv& C u_k v_k \; . 
\label{eq:psi} 
\ee
The radial shape of the Cooper pair wave function is deduced
from the Fourier transform of 
$u_k v_k=\Delta_n/2E_n(k)$ in Eq.~(\ref{eq:psi}).
The rms radius of Cooper pairs is then given by 
$\xi_{rms}=\sqrt{\langle r^2\rangle}=\sqrt{\int d r r^4 |\Psi_{pair}(r)|^2}$.
The rms radius $\xi_{rms}$ and the Pippard's coherence length,
$\xi_P=\hbar^2k_{Fn}/m^*_n\pi\Delta_n$, give similar 
size of the Cooper pair in the weak coupling regime.
The rms radius $\xi_{rms}$ is nevertheless a more appropriate quantity
in the BCS-BEC crossover region as well as in the strong BEC coupling region. 
In order to estimate the size of Cooper pairs a reference scale
is given by the average distance between neutrons $d_n$. 
If the rms radius of Cooper pairs is larger than $d_n$, the pair 
is interpreted as an extended BCS pair
while 
the Cooper pair will be considered as a compact BEC pair
if the rms radius is smaller than the average distance. 
Let us introduce another important quantity which also gives a measure of
the spatial correlations: the probability $P(r)$ for the partners of
the neutron Cooper pair to come close to each other within the
relative distance $r$, 
\be
P(r) = \int_0^r dr' r'^2 |\Psi_{pair}(r')|^2 \; .
\ee
The order parameters listed in Table~\ref{tab5} are closely
related. For instance, approximating $\xi_{rms}$ by $\xi_P$,
it could easily be shown that the ratio
$\xi_{rms}/d_n$ is proportional to $\epsilon_{Fn}/\Delta_n$.
Then, the strong coupling regime is reached 
if the ratio $\Delta_n/\epsilon_{Fn}$ is large.
The 
order parameter $\nu_n$,
the effective neutron chemical potential, could be interpreted as 
a half of the binding energy of Cooper pairs at finite density according
to the Schr\"odinger-like Eq.~(\ref{eq:sch}).

As shown in the Appendix~\ref{app:wf} it is convenient to decompose
the Cooper pair wave function into 
\be
\Psi_{pair}(r)=\Psi_1(r)+\Psi_2(r) \; ,
\label{eq:wf}
\ee
where
\be
\Psi_1(r)&=&C'\int_0^{k_\infty} dk \frac{k^2}{E_n(k)}\frac{\sin(kr)}{kr} 
\; , \label{eq:wf1} \\
\Psi_2(r)&=&C'\int_{k_\infty}^\infty dk \frac{k^2}{E_n(k)}\frac{\sin(kr)}{kr}
\label{eq:wf2} \; .
\ee
Choosing $k_\infty/k_0\gg 1$, with $k_0=\sqrt{2m^*_n\nu_n}/\hbar$,
it is possible to find an analytic 
expression for $\Psi_2$:
\be
\Psi_2(r)=-\frac{C'}{r} \frac{2 m^*_n}{\hbar^2} \,\si(k_\infty r)
\; , \label{eq:psi2}
\ee
where $\si(u)$ is the sinus integral defined as
$\si(u)=\int_u^\infty dz \,[\sin(z)/z]$.
It is clear from Eq.~(\ref{eq:psi2}) that the term $\Psi_2$ has a
$1/r$-type singularity. 
This singularity is due to the nature of
the contact interaction which does not contain a hard core repulsion.
With the hard core repulsion, the wave function goes to zero as
$r\rightarrow 0$~\cite{mat06}.
In the outer region ($r>3$~fm), the wave function behaves in the same way
if the contact interaction is deduced properly from the microscopic
%G-matrix 
calculations.
We checked the convergence of the wave function~(\ref{eq:wf}) with
respect to the parameter $k_\infty$. 
We found that the convergence is reached with $k_\infty\approx 2k_0$ as
is is shown in Fig.~\ref{fig11} in the Appendix~\ref{app:wf}.
In Ref.~\cite{mat06}, Matsuo introduced the cutoff momentum $k_c$ to
calculate the pair wave function~(\ref{eq:wf}).
We have compared the pair wave function $\Psi_{pair}(r)$ with the
one obtained by Matsuo. The two wave functions give essentially the 
same results, except for the low density region.
In the worst case, the wave function of Matsuo's treatment increases the
rms radius by about 10\% compared to the one obtained by the wave
 function~(\ref{eq:wf}).

%\begin{widetext}
%\onecolumngrid
%%%%%%%%%%%%%
%%%%%%%%%%%%%
%%%%%%%%%%%%%
\begin{figure*}[htb]
\begin{center}
\includegraphics[scale=0.5]{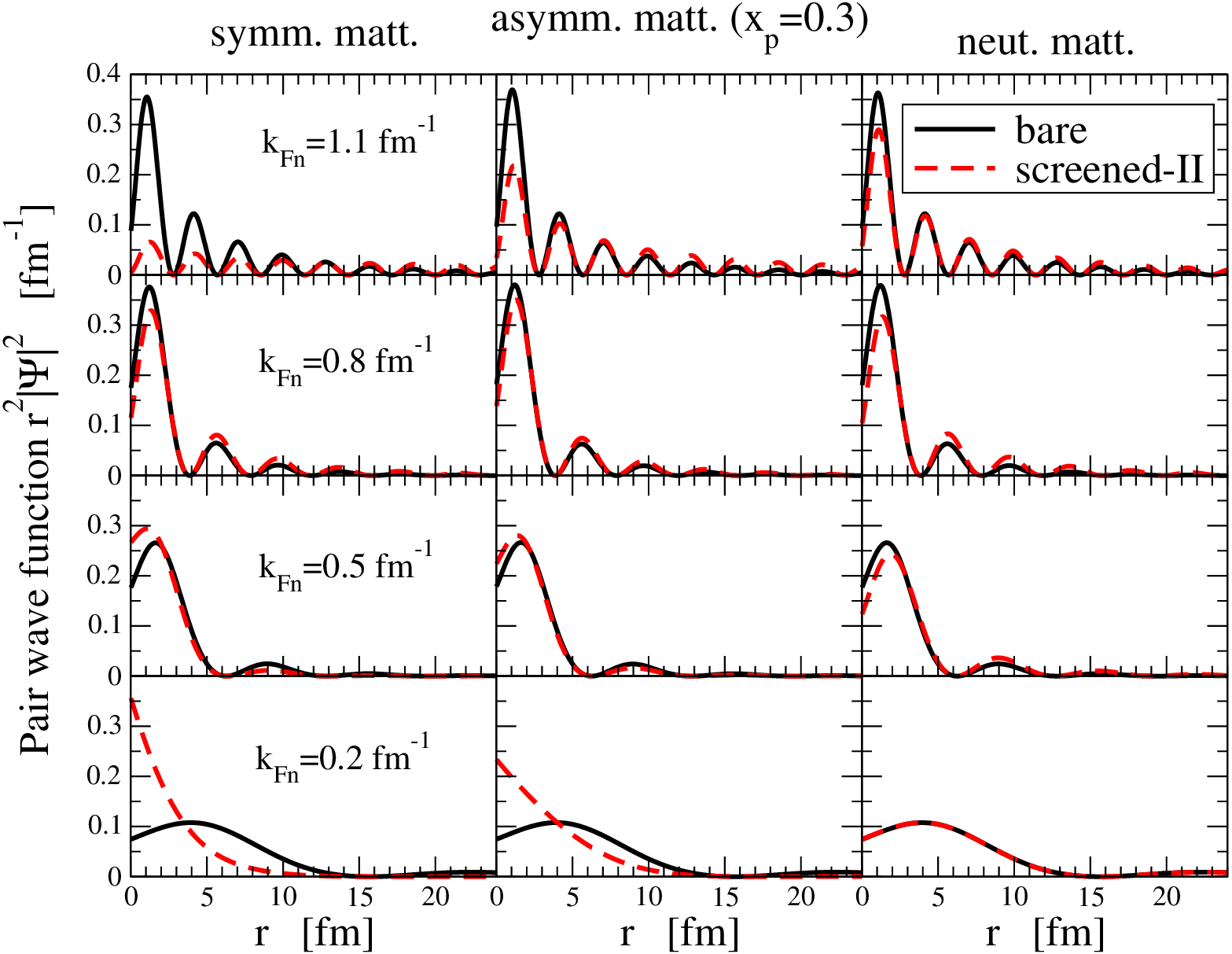}
\end{center}
\caption{(Color online) 
Neutron Cooper pair wave function $r^2|\Psi_{pair}(r)|^2$
as a function of the relative distance $r$ between the pair
partner at %densities corresponding to 
the Fermi momenta
$k_{Fn}$=1.1, 0.8, 0.5 and 0.2~fm$^{-1}$.}
\label{fig05}
\end{figure*}
%%%%%%%%%%%%%
% Obtenu avec gap.f
%%%%%%%%%%%%%
%%%%%%%%%%%%%
%\end{widetext}
%\twocolumngrid

The neutron Cooper pair wave function $r^2|\Psi_{pair}(r)|^2$ is shown
in Fig.~\ref{fig05} as a function of the relative distance $r$ between 
the pair partners taking different
Fermi momenta $k_{Fn}$=1.1, 0.8, 0.5 and
0.2~fm$^{-1}$, which correspond respectively to the 
densities: $\rho_n/\rho_0$=0.3, 0.1, 0.03 and 0.002. 
Calculations in symmetric, asymmetric and
neutron matter are shown
in the left, middle and right panels, respectively.
In Fig.~\ref{fig05}, 
we observe that the spatial extension and the profile of the
Cooper pair varies strongly with the density. 
A large extension is found close to the saturation density
at $k_{Fn}$=1.1~fm$^{-1}$. 
The profile of the wave function behaves 
as an oscillation convoluted by a decreasing exponent
and casts into the well known limit
$\sim K_0(r/\pi\xi_P)\sin(k_Fr)/k_Fr$~\cite{bcs57}. 
This indicates that the Cooper pair
is in the weak coupling BCS regime.
At lower densities, the Cooper pair shrinks and
the oscillation disappears.
The wave function resembles now the strong coupling limit (BEC)
$\sim\exp(-\sqrt{4m/\hbar^2 \; |\mu|}r)/r$~\cite{PS03}.
This is an indication that a possible BCS-BEC crossover may
occur in uniform matter.

%%%%%%%%%%%%%
%%%%%%%%%%%%%
%%%%%%%%%%%%%
\begin{figure}[b]
\begin{center}
\includegraphics[scale=0.36]{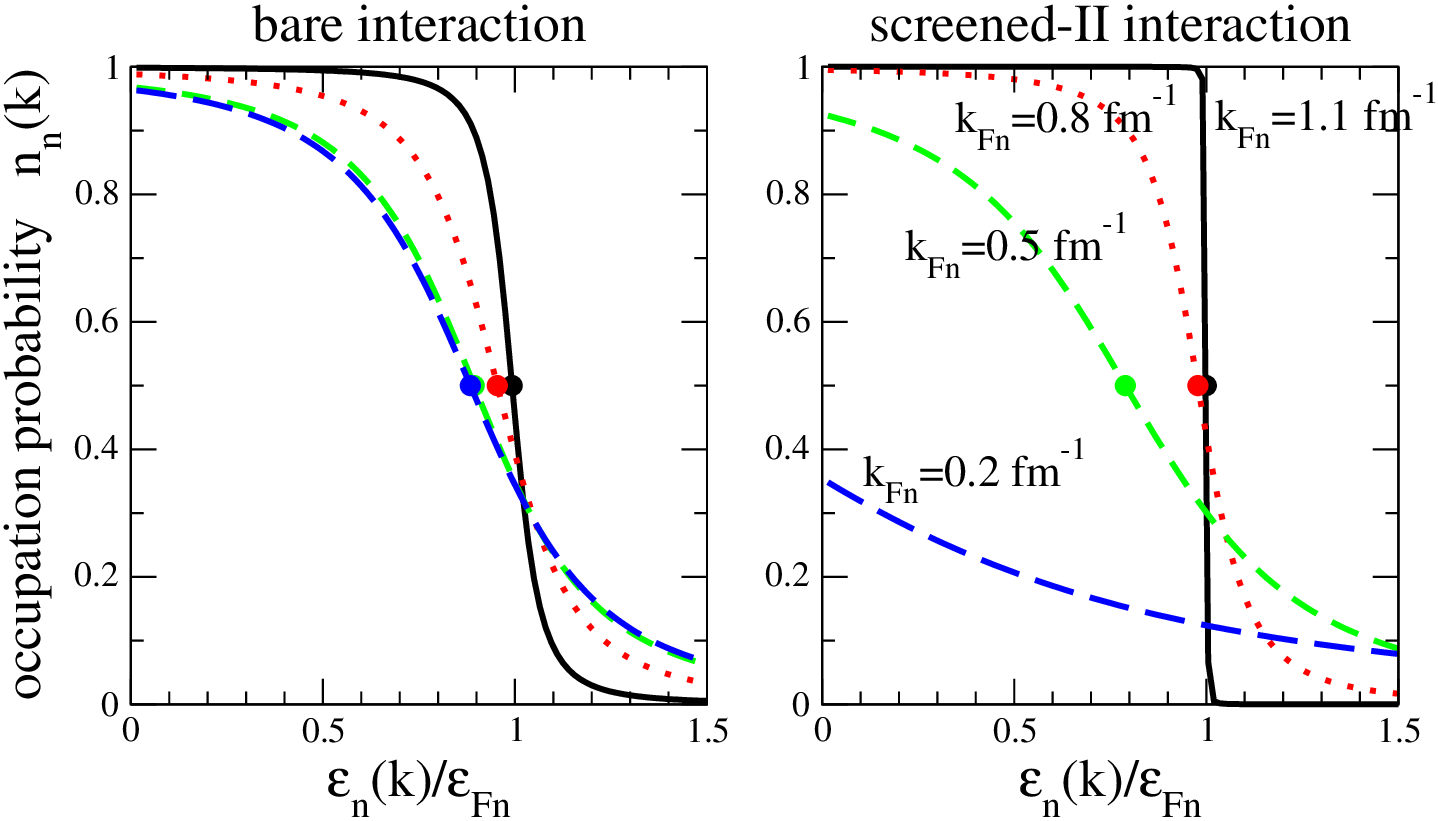}
\end{center}
\caption{(Color online) 
Occupation probability $n_n(k)$ for symmetric matter 
defined by Eq.~(\ref{eq:nn})
as a function of the ratio of the single particle kinetic energy 
to the Fermi energy $\epsilon_n(k)/\epsilon_{Fn}$ for a set of Fermi momenta
$k_{Fn}$= 1.1, 0.8, 0.5 and 0.2 fm$^{-1}$. 
We compare the results of the bare interaction (left panel) with the 
screened-II interaction (right panel). The values of the effective
neutron chemical potential $\nu_n/\epsilon_{Fn}$ are indicated by
the filled circles.}
\label{fig06}
\end{figure}
%%%%%%%%%%%%%
% Obtenu avec gap.f
%%%%%%%%%%%%%
%%%%%%%%%%%%%
It should be remarked that the latter limit seems 
well pronounced in
symmetric matter with the screened gap (see the panel at the
left bottom corner of Fig.~\ref{fig05}).
We show in Fig.~\ref{fig06}
the evolution of the occupation probability~(\ref{eq:nn}) in symmetric matter
for the two pairing
interactions.
For the screened-II interaction, the pairing correlations becomes strong
at low densities as the occupation probability is considerably different from the 
step function. 
In the case of the bare interaction, the correlations are not so
strong
to change $n_n(k)$ drastically even at low densities.
It should be noticed that
this analysis is independent of the detailed structure
of the Cooper pair wave function.
This change of the occupation probability proves
that the behavior of the Cooper pair wave 
function  
is not an artifact induced by the
zero range behavior of the contact interaction
but indeed is physical. 
It is clear that low density symmetric nuclear matter is
much more correlated with the screened-II interaction 
than with the bare one.
This is also the case for the BCS-BEC crossover as will be
discussed below.

%%%%%%%%%%%%%
%%%%%%%%%%%%%
%%%%%%%%%%%%%
\begin{figure*}[t]
\begin{center}
\includegraphics[scale=0.5]{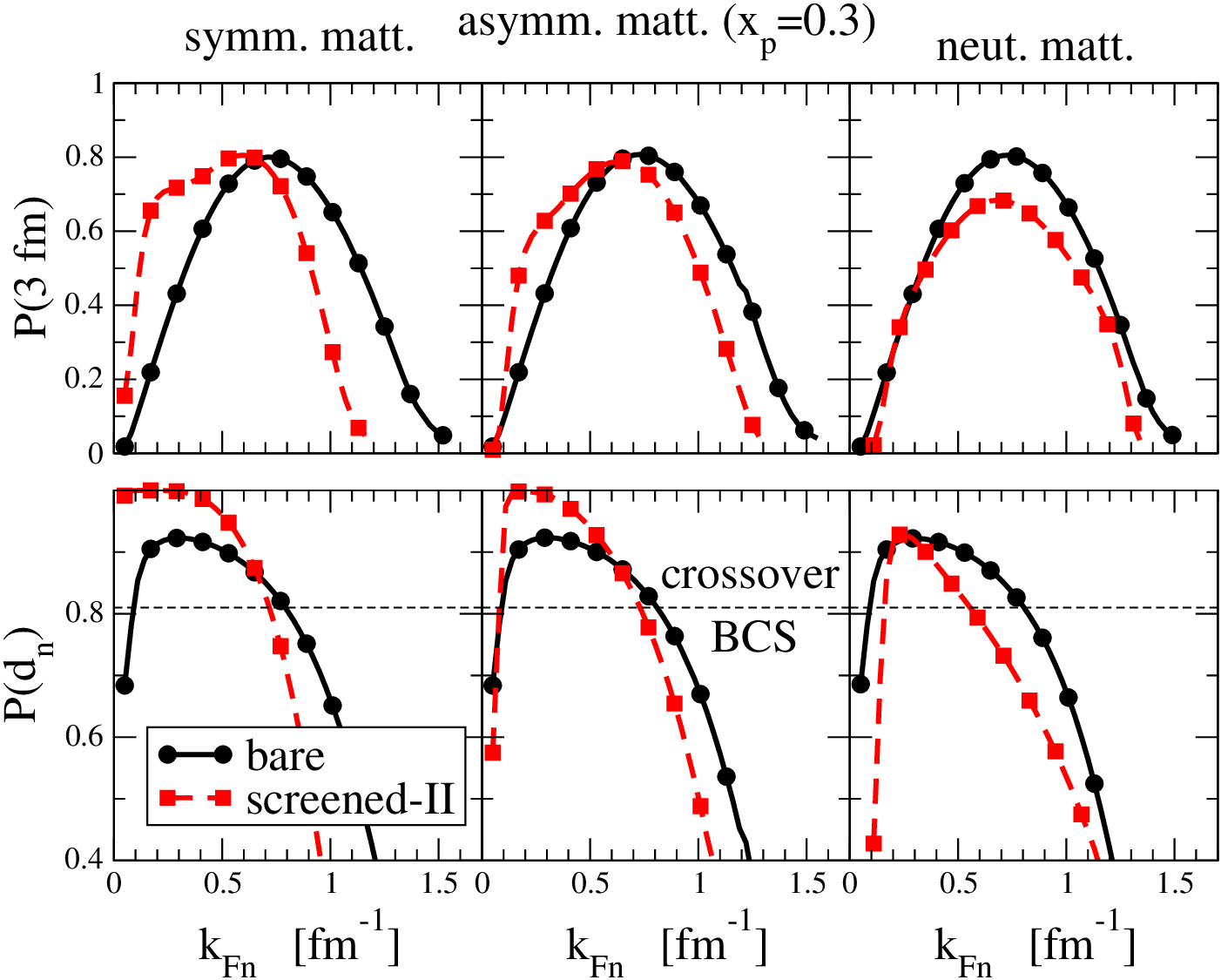}
\end{center}
\caption{(Color online) 
Probability $P$ for the partner neutrons within
 two typical lengths, 3 fm and $d_n$, as a function of the 
neutron Fermi momentum
$k_{Fn}$ in symmetric (left panel), asymmetric (central panel) and 
neutron matter (right panel).
The boundary of the BCS-BEC crossover is denoted by the dashed line.}
\label{fig07}
\end{figure*}
%%%%%%%%%%%%%
% Obtenu avec gap.f
%%%%%%%%%%%%%
%%%%%%%%%%%%%

Let us now discuss the BCS-BEC crossover which may depend
on the pairing interactions and also on the
asymmetry of the nuclear medium.
In the following, we study the different order parameters 
in Table~\ref{tab5} for the boundaries of the BCS-BEC 
phase transition.
Fig.~\ref{fig07} shows
the probabilities $P(r)$ for the partner neutrons to be correlated within
the typical scales, $r$=3 fm and $r$=$d_n$. The former scale is the typical 
range of the nucleon-nucleon force.
For the bare interaction, the probability $P(3~$fm$)$ 
has similar behavior in symmetric and asymmetric matter as a
  function of $k_{Fn}$.
For the screened-II interaction, there is a noticeable isospin dependence.
A low density shoulder appears in symmetric
matter, at around $k_{Fn}\sim 0.25$~fm$^{-1}$ ($\rho_n/\rho_0\sim 0.003$).
Then, it becomes smaller 
as the asymmetry increases and eventually disappears in neutron matter.
In neutron matter, the strong concentration of the pair wave
  function within the
interaction range 3~fm, $P(3$~fm$)>0.5$, is realized in the density region 
$k_{Fn}\sim 0.3-1.1$~fm$^{-1}$ (or $\rho_n/\rho_0\sim 0.007-0.3$) for
both 
pairing interactions. 
For symmetric matter, on the other hand, this region is
different for the two pairing interactions: 
the strong correlation occurs at much lower density region
for the screened-II interaction than for the bare one. 
This property can also be confirmed by the probability $P(d_n)$.
For the two pairing interactions, the Cooper pairs in
symmetric and asymmetric matter 
enter into the crossover regime at almost
the same density. 
The crossover in neutron matter occurs somewhat at lower density 
for the screened-II interaction.
As the density 
decreases, a different behavior is observed between the two pairing 
interactions for symmetric matter.
While the probability $P(d_n)$ decreases and goes back to the 
weak BCS regime for the bare interaction at very small density below
$k_{Fn}\sim 0.1$~fm$^{-1}$, the probability $P(d_n)$ 
continue to increase up to 1 for the screened-II interaction at very
low densities, $k_{Fn}<0.7$~fm$^{-1}$ ($\rho=n/\rho=0<0.07$~fm$^{-1}$).

%%%%%%%%%%%%%
%%%%%%%%%%%%%
%%%%%%%%%%%%%
\begin{figure*}[p]
\begin{center}
\includegraphics[scale=0.48]{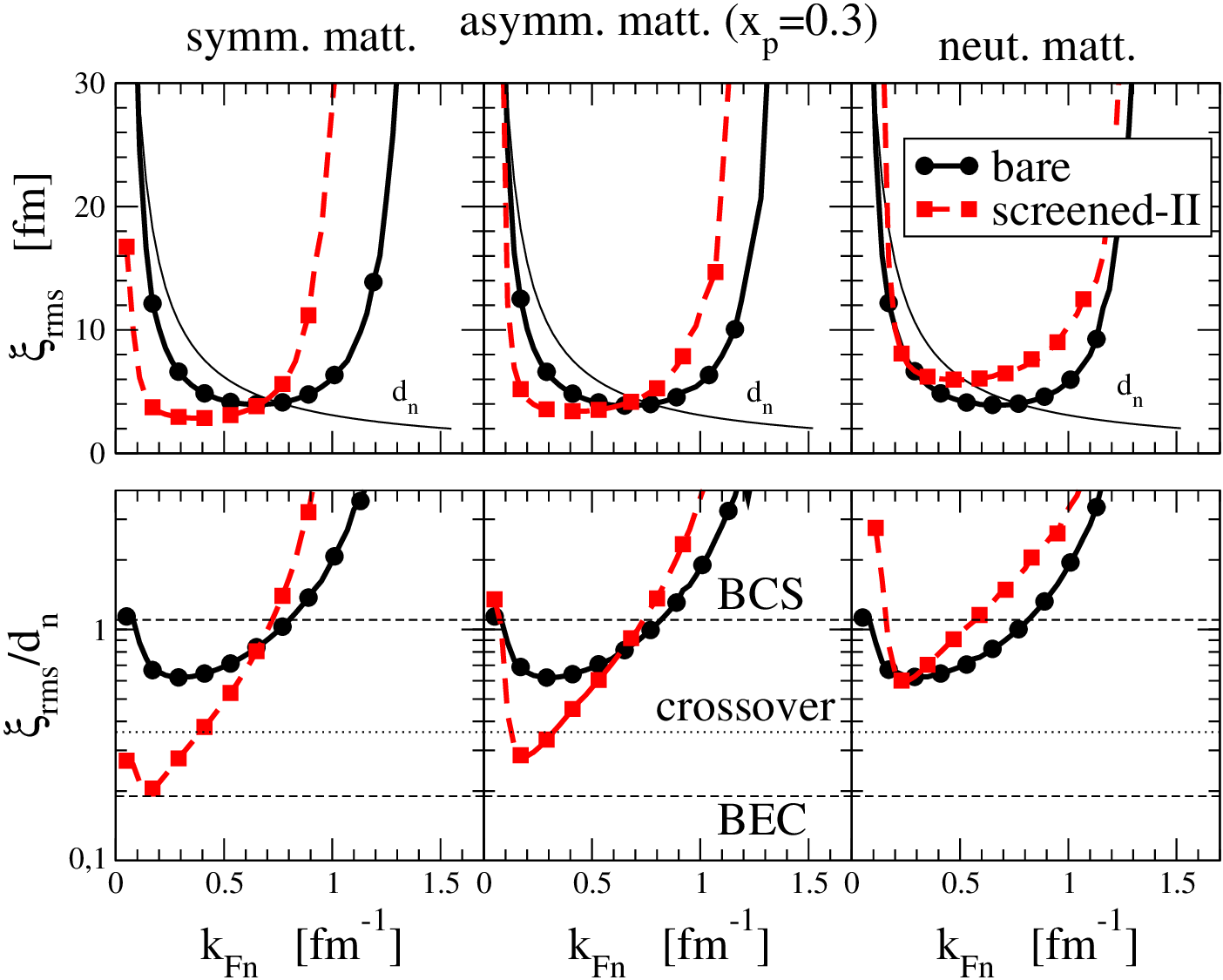}
\end{center}
\caption{(Color online) Top panels:
Comparison between the rms radius $\xi_{rms}$ of 
the neutron pair 
and the average inter-neutron
distance $d_n=\rho_n^{-1/3}$ (thin line) as a function of the neutron
Fermi momentum $k_{Fn}$ in symmetric (left panel), asymmetric
(central panel) and neutron matter (right panel).
Bottom panels: The order parameter $\xi_{rms}/d_n$ as a function of
$k_{Fn}$.
The boundaries of the BCS-BEC crossover are represented by
the two dashed lines, while the unitary limit is shown by the dotted line.
The two pairing interactions are used for the calculations.}
\label{fig08}
\end{figure*}
%%%%%%%%%%%%%
% Obtenu avec gap.f
%%%%%%%%%%%%%
%%%%%%%%%%%%%

%%%%%%%%%%%%%
%%%%%%%%%%%%%
%%%%%%%%%%%%%
\begin{figure*}[p]
\begin{center}
\includegraphics[scale=0.48]{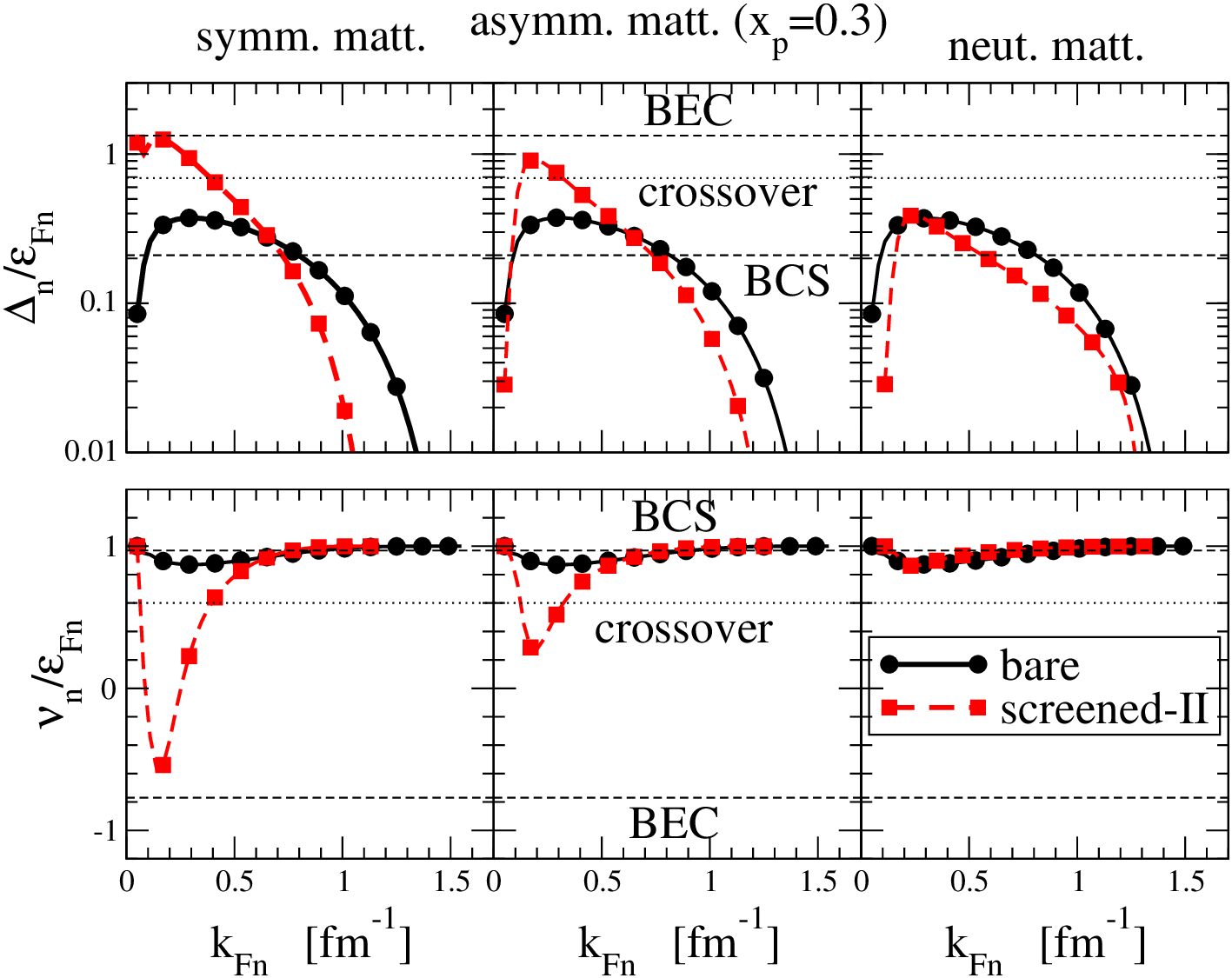}
\end{center}
\caption{(Color online) 
Ratios $\Delta_n/\epsilon_{Fn}$ and $\nu_n/\epsilon_{Fn}$ plotted 
as a function of the neutron Fermi momentum $k_{Fn}$ in symmetric 
(left panel), asymmetric (central panel) and neutron matter
(right panel).
The boundaries of the BCS-BEC crossover are shown by the 
two dashed lines, while the unitary limit is given by the dotted
line. See the text for details.} 
\label{fig09}
\end{figure*}
%%%%%%%%%%%%%
% Obtenu avec gap.f
%%%%%%%%%%%%%
%%%%%%%%%%%%%

We study further the BCS-BEC crossover by looking at the rms 
radius $\xi_{rms}$ and the neutron pairing gap $\Delta_n$.
In Fig.~\ref{fig08}, we show the rms radius $\xi_{rms}$ as a function 
of the
neutron Fermi momentum $k_{Fn}$ as well as the order parameter
$\xi_{rms}/d_n$. 
The rms radius of the Cooper pair is less than 5~fm in the region
$k_{Fn}\sim(0.4-0.9)$~fm$^{-1}$ ($\rho_n/\rho_0\sim 0.01-0.15$)
in the three panels for the bare interaction.
The screened-II interaction gives different effects in symmetric
and asymmetric matter: it increases the rms radius for the neutron matter,
while the rms radius stays small around 4~fm even at very low density
at $k_{Fn}\sim 0.15$~fm$^{-1}$ ($\rho_n/\rho_0\sim 0.0007$)
in symmetric matter.
In the lower panels is shown the ratio of the rms radius to the
average distance between neutrons $d_n$.
For the bare interaction, the size of the Cooper pair becomes smaller
than $d_n$ for the Fermi momentum $k_{Fn}<0.8$~fm$^{-1}$
($\rho_n/\rho_0\sim 0.1$) in general.
There are substantial differences for symmetric and asymmetric matter
in the case of the screened-II interaction. The crossover region
becomes smaller for the neutron matter, while the crossover region
increases in the cases of asymmetric ($x_p=0.3$) and symmetric
matter.
Especially, the correlations becomes strong in symmetric matter
and the Cooper pair reaches almost the BEC boundary at 
$k_{Fn}\sim 0.2$~fm$^{-1}$ ($\rho_n/\rho_0\sim 0.002$).
Notice that the two neutrons system is known experimentally to have a
virtual state in the zero density limit.
We have shown that this virtual state could lead to a strongly
correlated BEC state at low density in symmetric nuclear matter
according to the screened-II interaction.

The two other order parameters $\Delta_n/\epsilon_{Fn}$ and
$\nu_n/\epsilon_{Fn}$ are shown in Fig.~\ref{fig09}.
These results confirm the BCS-BEC crossover behavior which is
  found in Fig.~\ref{fig08}.
Namely, in symmetric matter, the gap $\Delta_n$ is much enhanced
by the screened interaction in the low density region, while no
enhancement can be seen in neutron matter.
As expected, the effective chemical potential $\nu_n$ induced by the
screened-II interaction becomes negative for
$k_{Fn}\sim 0.05-0.3$~fm$^{-1}$ ($\rho_n/\rho_0\sim 0.00002-0.01$)
in symmetric nuclear matter.
This strong correlation is reduced in asymmetric matter and is almost 
absent in the neutron matter as can be seen in Fig.~\ref{fig09}.
It should also be remarked that the order parameter
$\nu_n/\epsilon_{Fn}$, due to the Schr\"odinger-like Eq.~(\ref{eq:sch}), 
gives the same BCS-BEC crossover behavior as that of the order parameter
$\xi_{rms}/d_n$. 
The neutron effective chemical potential is then a good criteria
to discuss the BCS-BEC crossover.

%%%%%%%%%%%%%%%%%%%%%%%%%%%%%%%%%%%%%%%%%%
%%%%%%%%%%%%%%%%%%%%%%%%%%%%%%%%%%%%%%%%%%

\section{Conclusions}
\label{sec:conclusions}

%%%%%%%%%%%%%%%%%%%%%%%%%%%%%%%%%%%%%%%%%%
%%%%%%%%%%%%%%%%%%%%%%%%%%%%%%%%%%%%%%%%%%

A new type of density-dependent contact pairing interaction 
was obtained to reproduce
the pairing gaps in symmetric and neutron matter
obtained from a microscopic calculation~\cite{cao06}.
%calculated by the Brueckner G-matrix theory~\cite{cao06}.
The contact interactions reproduce the two types of
pairing gaps, i.e., the gap calculated with the bare
interaction and the gap modified by medium polarization effects.
It is shown that the medium polarization effects cannot be cast into the usual
density power law form in symmetric nuclear matter so that another
new isoscalar term $g_2$ in Eq.~(\ref{eq:g2}) is then added to the density
dependent term of the pairing interaction in 
Eq.~(\ref{eq:pairing_interaction}).

We have applied these density-dependent pairing interactions to
the study of the BCS-BEC crossover phenomenon in symmetric and
asymmetric nuclear matter.
We found that the spatial di-neutron correlation
is strong in general in a wide range of low matter densities, up to
$k_{Fn}\sim 0.9$~fm$^{-1}$ ($\rho_n/\rho_0\sim 0.15$).
This result is independent of the pairing interaction, either
  bare or screened one, as well as of
the asymmetry of the uniform matter. 
Moreover, it is shown that the two pairing interactions mentioned above lead to 
different features for BCS-BEC phase transition in symmetric
nuclear matter. 
To clarify the difference,
we studied various order parameters, the correlation probability
  $P(d_n)$, the rms radius of the Cooper pair $\xi_{rms}$, 
the gap $\Delta_n$ and
  the effective chemical potential $\nu_n$, as a function of the Fermi
  momentum $k_{Fn}$, or equivalently as a function of the density. The
  screened interaction enhances the BCS-BEC crossover phenomena in
  symmetric matter, while the pairing correlations as well as the
  crossover phenomena are decreased in neutron
matter by the medium polarization effects. For the screened-II
interaction, the crossover reaches almost to the BEC phase at
$k_{Fn}\sim 0.2$~fm$^{-1}$ in symmetric matter. We should notice,
however, that the BEC state is very sensible to the asymmetry of the medium
and disappears in neutron matter.

\acknowledgments
We are grateful to M.~Matsuo, P.~Schuck and N.~Sandulescu for
interesting discussions during the completion of this work.
This work was supported by the Japanese
Ministry of Education, Culture, Sports, Science and Technology
by Grant-in-Aid for Scientific Research under
the program number 19740115.

%%%%%%%%%%%%%%%%%%%%%%%%%%%%%%%%%%%%%%%%%%
%%%%%%%%%%%%%%%%%%%%%%%%%%%%%%%%%%%%%%%%%%

\appendix

%%%%%%%%%%%%%%%%%%%%%%%%%%%%%%%%%%%%%%%%%%
%%%%%%%%%%%%%%%%%%%%%%%%%%%%%%%%%%%%%%%%%%

\section{Effects of the cutoff prescription on pairing gap}
\label{app:co}

In the gap (\ref{eq:gap}), the integral runs over the 
momentum $k$, which is limited by the cutoff momenta $k_c^\pm$ to
avoid the ultraviolet divergence. 
There are several prescriptions for the cutoff
momenta depending on the physical problem for which the interaction is
applied.
\begin{description}
\item[Prescription 1:] This is the most simple prescription imposing on
the single particle kinetic energy with the condition
$\epsilon_n(k)<E_c$, i.e., $k_c^+ = \sqrt{2m^* E_c}/\hbar$ and $k_c^-=0$. 
It is independent of the Fermi momentum and of the pairing gap, but
still has a weak dependence on the density through 
the effective mass $m^*(\rho)$.
It has been used by several authors~\cite{ber91,esb92,esb97} and
also adopted in shell model calculations in which
all the shells up to a given cutoff energy are involved.
\item[Prescription 2:] This prescription is based on the fact that the
pairing occurs among states around the Fermi energy. Then, it is natural
to define the cutoff energy with respect to the Fermi momentum
by the condition $\epsilon_n(k)<\epsilon_{Fn}+E_c$, i.e.,
$k_c^+ = \sqrt{2m^* (\epsilon_{Fn}+E_c)}/\hbar$
and still $k_c^-=0$~\cite{mat06}.
Through $\epsilon_F$, the dependence on the density of this cutoff
is much stronger than for the prescription 1.
This prescription is close to the prescription used in HFB calculations. 

\item[Prescription 3:] This prescription is often used in
the HFB calculations for which the cutoff is defined with respect to the
quasi-particle energy 
$\sqrt{(\epsilon_n(k)-\nu_n)^2+\Delta_n^2}<E_c$.
This leads to the following definition of the cutoff momenta:
$k_c^\pm = \left[2m^*\left(\nu_n\pm\sqrt{E_c^2-\Delta^2}\right)\right]^{1/2}/\hbar$
(if $E_c>\Delta_n$).
If $k_c^-$ becomes
imaginary for very small $\nu_n$,
we set $k_c^-=0$.
In this prescription, the density dependence of the cutoff momenta
$k_c^\pm$ is not trivial since it depends on the chemical potential
and on the pairing gap. 
\end{description}

%%%%%%%%%%%%%
%%%%%%%%%%%%%
%%%%%%%%%%%%%
\begin{figure}[tb]
\begin{center}
\includegraphics[scale=0.36]{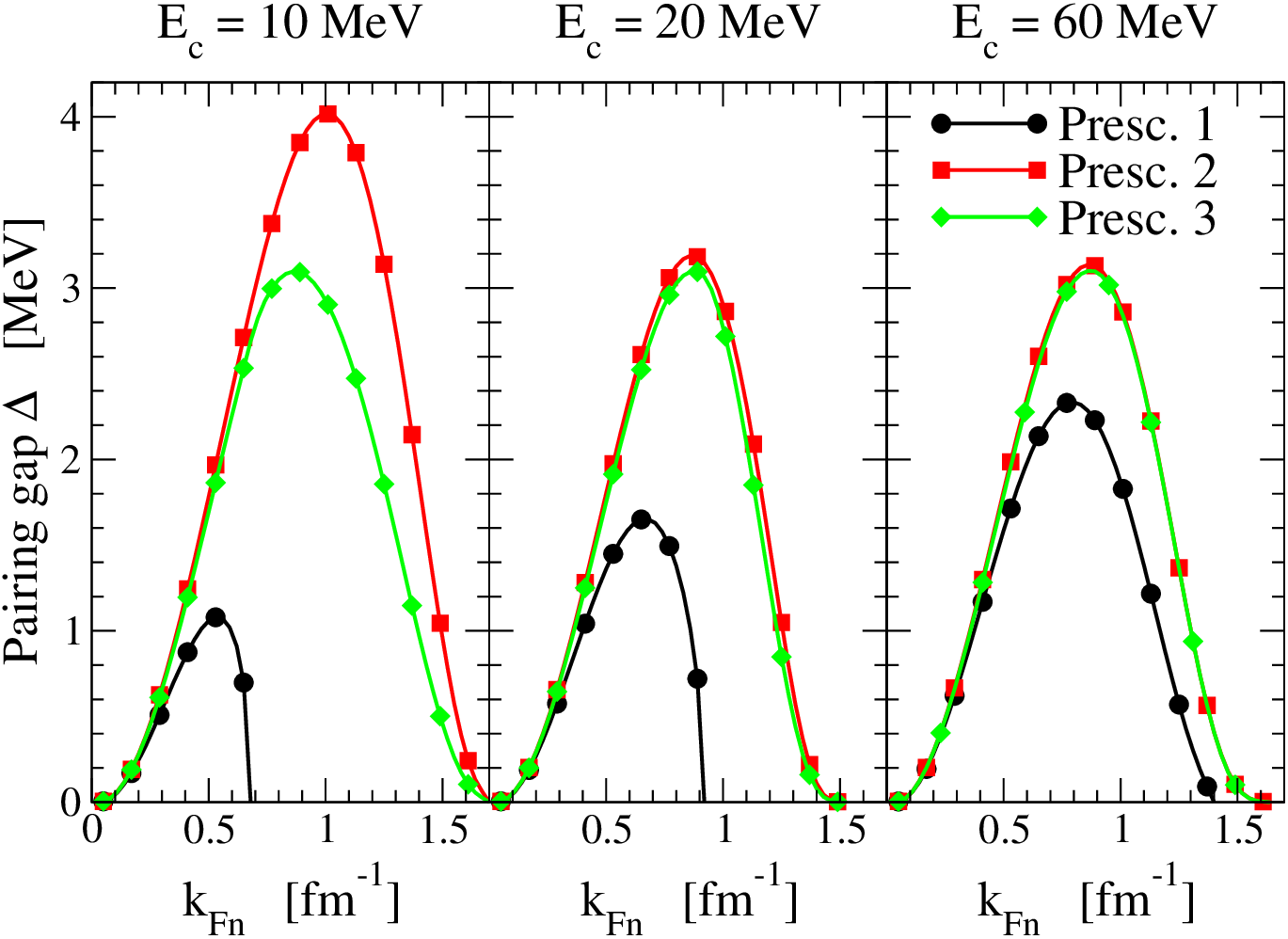}
\end{center}
\caption{(Color online) 
The pairing gap of symmetric nuclear
  matter as a function of the neutron Fermi momentum $k_{Fn}$
  for the 3 prescriptions. The solid, dashed, and dotted lines are for
  the prescriptions 1, 2, and 3, respectively. 
  Each panel corresponds to a fixed value for the 
cutoff energy, $E_c$=10, 20 or 60~MeV.}
\label{fig10}
\end{figure}
%%%%%%%%%%%%%
% Obtenu avec gap.f
%%%%%%%%%%%%%
%%%%%%%%%%%%%

It should be noted that in the limit $E_c\gg\Delta_n$, the
prescription~2 and 3 give the same cutoff momenta $k_c^\pm$.
If the limit $E_c\gg\epsilon_{Fn}$ is also satisfied, then
the three prescriptions are equivalent to each other.
In Fig.~\ref{fig10} are represented the pairing gaps obtained in
symmetric nuclear matter with the bare pairing interactions 
given in Table~\ref{tab2}.
We compare the pairing gaps calculated for the three prescriptions
with different values of the cutoff energy,
10, 20 and 60~MeV.
For a low value of the cutoff energy $E_c$=10~MeV, the three
prescriptions lead to very different pairing gaps.
For the cutoff energy larger than 20~MeV, the
prescriptions~2 and 3 give very similar results because
$E_c/\Delta_n\gg$1.
Notice that at very low Fermi momentum,
$k_{Fn}<0.4$~fm$^{-1}$ ($\rho_n/\rho_0<0.01$),
the three prescriptions give a similar pairing gap,
because both conditions $E_c\gg\Delta_n$ and 
$E_c\gg\epsilon_{Fn}$ are satisfied.
Nevertheless, for $k_{Fn}>0.4$~fm$^{-1}$,
the prescription~1 gives a pairing gap different from the
prescriptions~2 and 3, even for the larger cutoff energy $E_c=60$~MeV.
The reason is that the limit $E_c\gg\epsilon_{Fn}$ is not reached
for $k_{Fn}>0.4$~fm$^{-1}$.

%%%%%%%%%%%%%%%%%%%%%%%%
\begin{table}[b]
\begin{center}
\setlength{\tabcolsep}{.15in}
\renewcommand{\arraystretch}{1.3}
\begin{tabular}{ccccc}
\toprule 
             & \multicolumn{2}{c}{Presc.~1} & \multicolumn{2}{c}{Presc.~2}\\
$E_c$ [MeV] & $\eta_s$ & $\alpha_s$ & $\eta_s$ & $\alpha_s$ \\
\colrule
 60 & 0.461 & 0.579 & 0.593 & 0.537 \\
 40 & 0.413 & 0.487 & 0.657 & 0.506 \\
\botrule 
\end{tabular}
\end{center}
\caption{Parameters of the density-dependent term $g=g_1$ 
in Eq.~(\ref{eq:pairing_interaction}) obtained from the
fit to the bare pairing gap in symmetric nuclear matter.
The effective mass is obtained from SLy4 Skyrme interaction.
The parameters for the prescription~3 are already shown in
Table~\ref{tab2}.}
\label{tab3}
\end{table}%
%%%%%%%%%%%%%
% Obtenu avec gap.f
%%%%%%%%%%%%%
%%%%%%%%%%%%%

In Table~\ref{tab3} we give the parameters of the density-dependent
term $g=g_1$ of Eq.~(\ref{eq:pairing_interaction})
to fit the bare gap using the prescriptions~1 and 2
for symmetric nuclear matter.
Those for the prescription 3 are already given in
Table~\ref{tab2}. 
It shows how sensible these parameters are on the cutoff prescription.
One could remark that the parameter $\eta_s$ is much more affected 
by the prescription than the parameter $\alpha_s$.
We can also compare our results with other calculations.
The parameters obtained with the prescription~1 can be compared to
the one proposed in Ref.~\cite{esb97}, 
namely $\eta_s=0.45$, $\alpha_s=0.47$ for $E_c=60$~MeV.
The value of the parameter $\alpha_s$ is significantly different.
One can nevertheless obtain a comparable value for the parameter
$\alpha_s$ if one takes the
approximation: 
$\nu_n\sim \epsilon_{Fn}$.
With the prescription~2, we obtain similar
%comparable 
parameters to those in Ref.~\cite{mat06}
in which $\eta_s=0.60-0.63$, $\alpha_s=0.55-0.58$ are obtained.
The small differences can be explained by a different effective mass
and different adopted pairing gap.

\section{Cooper pair wave function}
\label{app:wf}

\begin{figure*}[htb]
\begin{center}
\includegraphics[scale=0.5]{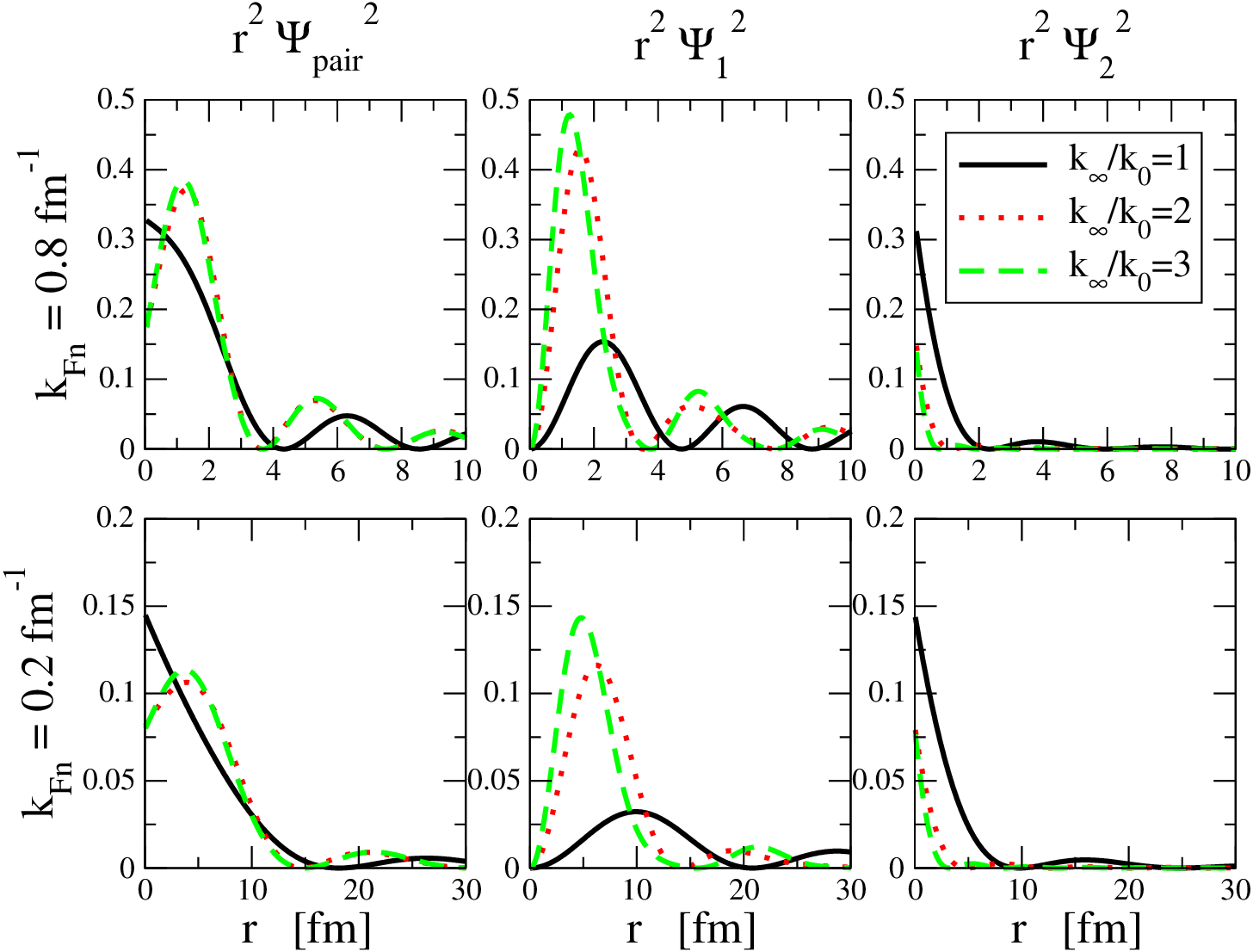}
\end{center}
\caption{(Color online) The Cooper pair wave function $r^2\Psi_{pair}^2$
as well as its individual contribution $r^2\Psi_1^2$ and $r^2\Psi_2^2$
for different values of the ratio $k_\infty/k_0$ and for two values of the
neutron Fermi momentum. Notice that the convergence 
with respect to the ratio $k_\infty/k_0$ is very fast.}
\label{fig11}
\end{figure*}

From the Cooper pair wave function~(\ref{eq:psi}) in the momentum
  space,
%and using $u_k v_k=\Delta_n/E_n(k)$, 
we obtain the radial dependence of the Cooper pair wave function
  by a Fourier transform,
\be
\Psi_{pair}(r) &=& \frac{C}{(2\pi)^3} \int d^3k \; u_k v_k e^{i\vec
  k \cdot \vec r} \\
 &=& C' \int dk \frac{k^2}{E_n(k)} \frac{\sin kr} {kr} 
\ee
where the normalization constant $C'$ is determined from the condition 
$\int dr r^2 |\Psi_{pair}(r)|^2=1$.
The wave function $\Psi_{pair}$ is separated into two terms 
$\Psi_1$ and $\Psi_2$ 
defined in Eqs.~(\ref{eq:wf1})-(\ref{eq:wf2}).
The term $\Psi_1$ is solved numerically.
If $k_\infty/k_0\gg 1$, with $k_0=\sqrt{2m^*_n\nu_n}/\hbar$, 
we obtain the analytical form~(\ref{eq:psi2}) for $\Psi_2(r)$
to the first order in $\nu_n/\epsilon_n(k)$, 
$1/E_n(k)\sim 2m^*_n/\hbar^2 k^2$.
It is shown in Fig.~\ref{fig11} that the convergence is very
fast and $k_\infty=2k_0$ provides already a good converged solution.
%Note however, that the pair wave function $\Psi_{pair}$
%  represented on the left panels of Fig.~\ref{fig11} converge much
%  faster than the functions $\Psi_1$ and $\Psi_2$ individualy.

%%%%%%%%%%%%%
%%%%%%%%%%%%%
%%%%%%%%%%%%%
%%%%%%%%%%%%%
% Obtenu avec gap.f
%%%%%%%%%%%%%
%%%%%%%%%%%%%

From the decomposition into $\Psi_1$ and $\Psi_2$ presented in
Eq.~(\ref{eq:wf}), the treatment of Matsuo is
equivalent to set $k_\infty=k_c$ and $\Psi_2=0$, where $k_c$ is
the cutoff momentum according to the prescription 2 in 
Appendix~\ref{app:co}.
By this treatment,
a good agreement between the wave function obtained with
the contact interaction and the one obtained with the Gogny interaction is
reached for a cutoff energy $E_c=30$~MeV. 
The agreement is very nice especially in the region $r<3$~fm.
This result is easily
understood from the decomposition of $\Psi$ into $\Psi_1$ and $\Psi_2$.
Namely, by introducing a cutoff energy in the definition of the
Cooper pair wave function, Matsuo's modification effectively removes
the singularity at $r\sim 0$. 
The justification of 
cutoff 
in the Cooper pair
wave function could be understood from model space considerations. 
For practical reasons, the calculations for finite systems are
never done with infinite basis but in a sub-basis involving a finite
number of wave functions. 
A cutoff is then naturally introduced in 
the finite model
space and all the quantities, including the Cooper pair wave function, 
are calculated 
within the same model space.
The cutoff treatment of Matsuo is thus common in nuclear matter
and finite nucleus calculations.
However, it should be remarked that despite the singularity of the
wave function in nuclear matter, 
the rms radius $\xi_{rms}$ stays finite and larger than 3~fm, in the
range of explored densities. 
We have checked that the treatment of Matsuo only affects the rms
radius by about 10\% in the worst situation, i.e.,
it increases the rms radius by less than 0.5~fm for neutron density
at $\rho_n\sim\rho_0/10$.

%%%%%%%%%%%%%%%%%%%%%%%%%%%%%%%%%%%%%%%%%%
%%%%%%%%%%%%%%%%%%%%%%%%%%%%%%%%%%%%%%%%%%


\begin{thebibliography}{99}
\bibitem{bal04} M. Baldo, C. Maieron, P. Schuck and X. Vi\~nas, 
Nucl.Phys. {\bf A736}, 241 (2004).
\bibitem{sed03} A. Sedrakian, T. T. S. Kuo, H. M\"uther, and
  P. Schuck, Phys. Lett. {\bf B576}, 68 (2003).
\bibitem{bro00} B.A. Brown, Phys. Rev. Lett {\bf 85}, 5296 (2000). 
\bibitem{bar06} V. Baran and J. Margueron, Eur. Phys. J. {\bf A 30}, 141 (2006); 
J. Margueron and Ph. Chomaz, Phys. Rev. {\bf C 71}, 024318 (2005); 
V. Baran et al., Phys. Rev. Lett. {\bf 86}, 4492 (2001). 
\bibitem{hor05} C.J. Horowitz, A. Schwenk,
Nucl.Phys. {\bf A776}, 55 (2006).
\bibitem{sed06} A. Sedrakian and J. W. Clark, Phys. Rev. {\bf C 73},
  035803 (2006).
\bibitem{bal95} M. Baldo, U. Lombardo, and P. Schuck, Phys. Rev. 
{\bf C 52}, 975 (1995).
\bibitem{mar07} J. Margueron, E. van Dalen, and C. Fuchs, Phys. Rev. 
{\bf C 76}, 034309 (2007).
\bibitem{bul05} A. Bulgac, Phys. Rev. Lett. {\bf 95}, 140403 (2005).
\bibitem{lom99} U. Lombardo, in "Nuclear Methods and the Nuclear Equation of State", 
Int. Rev. of Nucl. Physics, Vol. 9, M. Baldo Eds. (World-Scientific, Singapore, 1999).
\bibitem{cao06} L. G. Cao, U. Lombardo, and P. Schuck, Phys. Rev. {\bf C
  74}, 064301 (2006).
\bibitem{hei00} H. Heiselberg, C.J. Pethick, H. Smith, and L. Viverit,
  Phys. Rev. Lett. {\bf 85}, 2418 (2000).
\bibitem{sch02} H.-J. Schulze, A. Polls, and A. Ramos, Phys. Rev. {\bf
  C 63}, 044310 (2001).
\bibitem{ber91} G. F. Bertsch and H. Esbensen, Ann. Phys. (N.Y.) 209, 327 (1991).
\bibitem{esb92} H. Esbensen and G. F. Bertsch, Nucl. Phys. {\bf A542}, 310 (1992).
\bibitem{esb97} H. Esbensen, G. F. Bertsch and K. Hencken,
  Phys. Rev. {\bf C 56}, 3054 (1997).
\bibitem{dob96} J. Dobaczewski, W. Nazarewicz, T. R. Werner, J. F. 
Berger, C. R. Chinn, and J. Decharg\'e, 
Phys. Rev. {\bf C 53}, 2809 (1996).
\bibitem{hag05} K. Hagino and H. Sagawa, Phys. Rev. {\bf C 72}, 044321 (2005).
\bibitem{pil07} N. Pillet, N. Sandulescu, and P. Schuck, Phys. Rev. {\bf C 76}, 024310 (2007).
\bibitem{cho04} Ph. Chomaz, M. Colonna, J. Randrup, Phys. Rep. {\bf 389}, 263 (2004).
\bibitem{yak04} D. G. Yakovlev and C. J. Pethick, Annu. Rev. Astron. 
Astrophys. {\bf 42}, 169 (2004).
\bibitem{sed07} A. Sedrakian, Prog. Part. Nucl. Phys. {\bf 58}, 168 (2007).
\bibitem{mon07} Ch. Monrozeau, J. Margueron, and N. Sandulescu,
  Phys. Rev. {\bf C 75}, 065807 (2007). 
\bibitem{hag07} K. Hagino, H. Sagawa, J. Carbonell, and P. Schuck,
  Phys. Rev. Lett. {\bf 99}, 022506 (2007).
\bibitem{mat06} M. Matsuo, Phys. Rev. {\bf C 73}, 044309 (2006).
\bibitem{sch03} A. Schwenk, B. Friman, and G.E. Brown, Nucl. Phys. {\bf
  A713}, 191 (2003).
\bibitem{fab05}A. Fabrocini, S. Fantoni, A. Yu. Illarionov, and K. E. Schmidt,
Phys. Rev. Lett. {\bf 95}, 192501 (2005).
\bibitem{abe07} T. Abe, and R. Seki, arXiv:nucl-th/0708.2523.
\bibitem{gez07} A. Gezerlis, and J. Carlson, arXiv:nucl-th/0711.3006.
\bibitem{bar99} F. Barranco, R. A. Broglia, G. Gori, E. Vigezzi, P.-F. Bortignon and 
J. Terasaki, Phys. Rev. Lett. {\bf 83}, 2147 (1999).
\bibitem{gio02} N. Giovanardi, F. Barranco, R.A. Broglia, and
  E. Vigezzi, Phys. Rev. {\bf C 65}, 041304(R) (2002).
\bibitem{bar05} F. Barranco, P.F. Bortignon, R.A. Broglia, G. Col\`o,
  P. Schuck, E. Vigezzi, and X. Vi\~nas, Phys. Rev. {\bf C 72}, 054314 (2005).
\bibitem{gar99} E. Garrido, P. Sarriguren, E. Moya de Guerra, and
  P. Schuck, Phys. Rev. {\bf C 60}, 064312 (1999).
\bibitem{gor04} Goriely et al., Nucl. Phys. {\bf A739}, 331 (2004);
  Phys. Rev. {\bf C 66}, 024326 (2002); Phys. Rev. {\bf C 68}, 054325 (2003).
\bibitem{bcs57} J. Bardeen, L. N. Cooper, and J. R. Schrieffer,
Phys. Rev. {\bf 108}, 1175 (1957).
\bibitem{gennes} P. G. de Gennes, {\it Superconductivity of Metals and
  Alloys}, (Addison-Wesley, Reading, MA, 1989).
\bibitem{schuck} P. Ring and P. Schuck, The Nuclear Many-body Problem
(Springer-Verlag, Heidelberg, 1980).
\bibitem{noz85} Ph. Nozi\`eres and S. Schmitt-Rink, J. Low
  Temp. Phys. {\bf 59}, 195 (1985).
\bibitem{PS03}P. Pieri and G.C. Strinati, Phys. Rev. Lett., {\bf 91},
  030401 (2003).
\bibitem{eng97} J. R. Engelbrecht, M. Randeria, C. A. R. S\'{a} de
  Melo, Phys. Rev {\bf B 55}, 15153 (1997).
\bibitem{pap99} T. Papenbrock and G. F. Bertsch, Phys. Rev. {\bf C
  59}, 2052 (1999).
\end{thebibliography}
\end{document}